\documentclass[prx,twocolumn,superscriptaddress,citeautoscript,showpacs,amsart,longbibliography]{revtex4-2}

\usepackage{blindtext}
\usepackage{graphicx}
\usepackage{bm}
\usepackage{epsfig}
\usepackage{amssymb}
\usepackage{amsfonts}
\usepackage{braket}
\usepackage{color}
\usepackage{epstopdf}
\usepackage{hyperref}
\usepackage{float}
\restylefloat{table}
\usepackage{bibentry}
\usepackage{color}
\usepackage{multirow}
\usepackage[caption=false]{subfig}

\usepackage{amsfonts}
\usepackage{amsthm}
\usepackage{graphicx}
\usepackage{multirow}
\usepackage{color}
\usepackage{bbold}
\usepackage{bm}
\usepackage{times}
\usepackage{amsmath,bm,amsfonts}
\usepackage{dcolumn}
\usepackage{graphicx}
\usepackage{latexsym}

\newcommand{\bpm}{\begin{pmatrix}}
\newcommand{\epm}{\end{pmatrix}}
\newcommand{\ba}{\begin{eqnarray}}
\newcommand{\ea}{\end{eqnarray}}
\newcommand{\bd}{\begin{displaymath}}

\graphicspath{{figures/}}% Put all figures in this directory.

\begin{document}

\title{Heteroepitaxial control of Fermi liquid, Hund metal, and Mott insulator phases in the single-atomic-layer limit}

\author{Jeong Rae Kim}
\thanks {These authors contributed equally to this work.}
\affiliation{Center for Correlated Electron Systems, Institute for Basic Science, Seoul 08826, Korea}
\affiliation{Department of Physics and Astronomy, Seoul National University, Seoul 08826, Korea}

\author{Byungmin Sohn}
\thanks {These authors contributed equally to this work.}
\affiliation{Center for Correlated Electron Systems, Institute for Basic Science, Seoul 08826, Korea}
\affiliation{Department of Physics and Astronomy, Seoul National University, Seoul 08826, Korea}
\affiliation{Department of Applied Physics, Yale University, New Haven, Connecticut 06520, USA}

\author{Hyeong Jun Lee}
\thanks {These authors contributed equally to this work.}
\affiliation{Center for Theoretical Physics of Complex Systems, Institute for Basic Science (IBS), Daejeon 34126, Republic of Korea}

\author{Sangmin Lee}
\affiliation{Department of Materials Science and Engineering and Research Institute of Advanced Materials, Seoul National University, Seoul 08826, Korea}

\author{Eun Kyo Ko}
\affiliation{Center for Correlated Electron Systems, Institute for Basic Science, Seoul 08826, Korea}
\affiliation{Department of Physics and Astronomy, Seoul National University, Seoul 08826, Korea}

\author{Sungsoo Hahn}
\affiliation{Center for Correlated Electron Systems, Institute for Basic Science, Seoul 08826, Korea}
\affiliation{Department of Physics and Astronomy, Seoul National University, Seoul 08826, Korea}

\author{Sangjae Lee}
\affiliation{Center for Correlated Electron Systems, Institute for Basic Science, Seoul 08826, Korea}

\author{Younsik Kim}
\affiliation{Center for Correlated Electron Systems, Institute for Basic Science, Seoul 08826, Korea}
\affiliation{Department of Physics and Astronomy, Seoul National University, Seoul 08826, Korea}

\author{Donghan Kim}
\affiliation{Center for Correlated Electron Systems, Institute for Basic Science, Seoul 08826, Korea}
\affiliation{Department of Physics and Astronomy, Seoul National University, Seoul 08826, Korea}

\author{Hong Joon Kim}
\affiliation{Center for Correlated Electron Systems, Institute for Basic Science, Seoul 08826, Korea}
\affiliation{Department of Physics and Astronomy, Seoul National University, Seoul 08826, Korea}

\author{Youngdo Kim}
\affiliation{Center for Correlated Electron Systems, Institute for Basic Science, Seoul 08826, Korea}
\affiliation{Department of Physics and Astronomy, Seoul National University, Seoul 08826, Korea}

\author{Jaeseok Son}
\affiliation{Center for Correlated Electron Systems, Institute for Basic Science, Seoul 08826, Korea}
\affiliation{Department of Physics and Astronomy, Seoul National University, Seoul 08826, Korea}

\author{Charles H. Ahn}
\affiliation{Department of Applied Physics, Yale University, New Haven, Connecticut 06520, USA}
\affiliation{Department of Physics, Yale University, New Haven, Connecticut 06520, USA}

\author{Frederick J. Walker}
\affiliation{Department of Applied Physics, Yale University, New Haven, Connecticut 06520, USA}

\author{Ara Go}
\affiliation{Department of Physics, Chonnam National University, Gwangju 61186, Republic of Korea}

\author{Miyoung Kim}
\affiliation{Department of Materials Science and Engineering and Research Institute of Advanced Materials, Seoul National University, Seoul 08826, Korea}

\author{Choong H. Kim}
\email[Electronic address:$~~$]{chkim82@snu.ac.kr}
\affiliation{Center for Correlated Electron Systems, Institute for Basic Science, Seoul 08826, Korea}
\affiliation{Department of Physics and Astronomy, Seoul National University, Seoul 08826, Korea}

\author{Changyoung Kim}
\email[Electronic address:$~~$]{changyoung@snu.ac.kr}
\affiliation{Center for Correlated Electron Systems, Institute for Basic Science, Seoul 08826, Korea}
\affiliation{Department of Physics and Astronomy, Seoul National University, Seoul 08826, Korea}

\author{Tae Won Noh}
\email[Electronic address:$~~$]{twnoh@snu.ac.kr}
\affiliation{Center for Correlated Electron Systems, Institute for Basic Science, Seoul 08826, Korea}
\affiliation{Department of Physics and Astronomy, Seoul National University, Seoul 08826, Korea}

\date{\today}

\begin{abstract}

Interfaces between dissimilar correlated oxides can offer devices with versatile functionalities. In that respect, manipulating and measuring novel physical properties of oxide heterointerfaces are highly desired. Yet, despite extensive studies, obtaining direct information on their momentum-resolved electronic structure remains a great challenge. This is because most correlated interfacial phenomena appear within a few atomic layers from the interface, thus limiting the application of available experimental probes. Here, we utilize atomic-scale epitaxy and photoemission spectroscopy to demonstrate the interface control of correlated electronic phases in atomic-scale ruthenate--titanate heterostructures. While bulk SrRuO$_3$ is a ferromagnetic metal, the heterointerfaces exclusively realize three distinct correlated phases in the single-atomic-layer limit. Our theory reveals that atomic-scale structural proximity effects lead to the emergence of Fermi liquid, Hund metal, and Mott insulator phases in the quantum-confined SrRuO$_3$. These results highlight the extensive interfacial tunability of electronic phases, hitherto hidden in the atomically thin correlated heterostructure.

\end{abstract}
\maketitle

\section*{Introduction}

Artificial electronic states confined at heterointerfaces are a basis for modern semiconductor electronics and a fundamental theme in solid state physics~\cite{klitzing1980new,tsui1982two,qing2012interface,fu2008superconducting,sze2021physics}. While most devices thus far have utilized electrical charge modulation, the introduction of other physical degrees of freedom suggests an opportunity to design novel interfacial properties. In that context, strongly correlated electron systems of transition metal oxides provide optimal interfaces where charge, spin, orbital, and lattice degrees of freedom interact with each other, leading to novel properties~\cite{hwang2012emergent,zubko2011interface}. Notably, charge modulation at the correlated oxide interfaces is characterized by a significantly shorter length scale compared to that of semiconductor heterojunctions, as the charge screening in correlated electronic phases is usually expected to be of atomic-scale due to their high carrier densities~\cite{ahn2003electric,chakhalian2012whither,nelson2022interfacial}. Furthermore, previous studies have shown that magnetic~\cite{takahashi2001interface}, orbital~\cite{chakhalian2007orbital}, and structural~\cite{kan2016tuning} reconstructions also occur within several atomic layers from the interface. The short length scale of the interfacial reconstruction indicates the spatially abrupt modification of physical properties, which can translate into a high level of integration of tunable correlated oxide interfaces. 

Despite the great potential that oxide interfaces can offer, the atomic length scale required for the interface control constitutes the greatest challenge when conducting relevant experiments. The spatially confined nature of the interfacial electronic states requires the synthesis of atomically abrupt epitaxial oxide heterostructures. Even for such high-quality samples, transport measurements often suffer from severe interface scattering as well as percolation problems over the measurement length scale~\cite{sohn2021observation}, impeding observation of their intrinsic properties. Systematic transport measurements have been limited to a few high-electron-mobility oxide interfaces with carrier densities comparable to the densities of conventional semiconductors~\cite{kozuka2009two,falson2018review}. Therefore, an alternative experimental approach is desired for observation of the intrinsic properties and thus a comprehensive understanding of the correlated interfacial phenomena.

Here, we demonstrate that quantum confinement and structural proximity effects can be combined to realize a wide range of correlated electronic phases at epitaxial oxide heterointerfaces in the single-atomic-layer limit. Using various heteroepitaxial buffer layers, we selectively manipulated the oxygen octahedral arrangement of SrRuO$_3$ (SRO) layers. From the bulk ferromagnetic metal, our interface engineering gives three distinct correlated electronic phases in the single-atomic-layer limit. We theoretically investigate those phases and classify them as Fermi liquid, Hund metal, and Mott insulator.

%%%%%%%%%%%%%%%%%%%%%%%%%%%%%%%%%%%%%%%%%%%%%%%%%%%%%%%%%%%%%%%
\begin{figure}[]
\includegraphics[width=\linewidth]{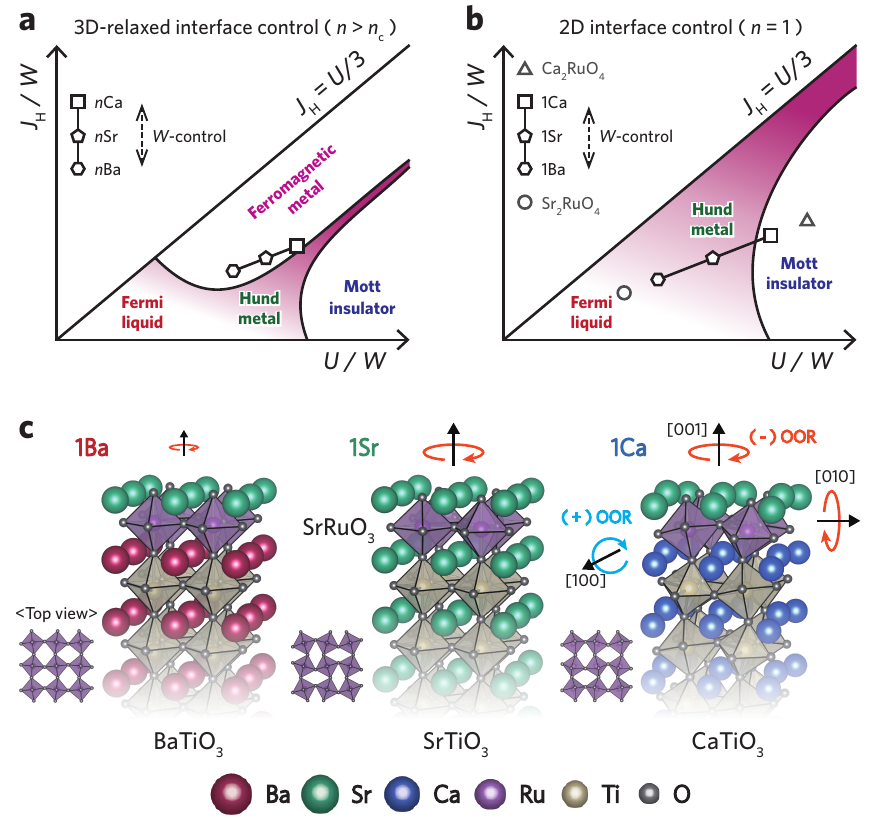}
\centering
\caption{{\bf Interface control of correlated electronic phases in epitaxial SrRuO$_3$ (SRO) heterostructures.} 
         {\bf a}, Schematic of the electronic phase diagram of thick SRO films, wherein the bandwidth ({\it W}) varies. Coulomb interaction ({\it U}) and Hund coupling ({\it J}$_H$) are almost intrinsic constants for the Ru atom. Above the critical thickness ({\it n}$_c$), the effect of interface control on the {\it n}-unit-cell ({\it n}-UC) SRO film is three-dimensionally relaxed as it progresses through multiple atomic layers.
         {\bf b}, Schematic of the electronic phase diagram of the single-atomic-layer ruthenates, wherein the {\it W} varies. Through quantum confinement effects, ferromagnetic metal phases are suppressed. The effect of interface control is maximal in the single-atomic-layer limit.
         {\bf c}, Schematic illustrations of the atomic configuration of the monolayer SROs interfaced with ATiO$_3$ (1A, A $=$ Ba, Sr, and Ca). The out-of-plane-oriented oxygen octahedral rotation (OOR) for 1Ba (1Sr) is small (large). 1Ca has both in-plane-oriented and out-of-plane-oriented OOR.
         }
\label{fig:1}
\end{figure}
%%%%%%%%%%%%%%%%%%%%%%%%%%%%%%%%%%%%%%%%%%%%%%%%%%%%%%%%%%%%%%

%%%%%%%%%%%%%%%%%%%%%%%%%%%%%%%%%%%%%%%%%%%%%%%%%%%%%%%%%%%%%%%
\begin{figure*}[]
\includegraphics[width=\linewidth]{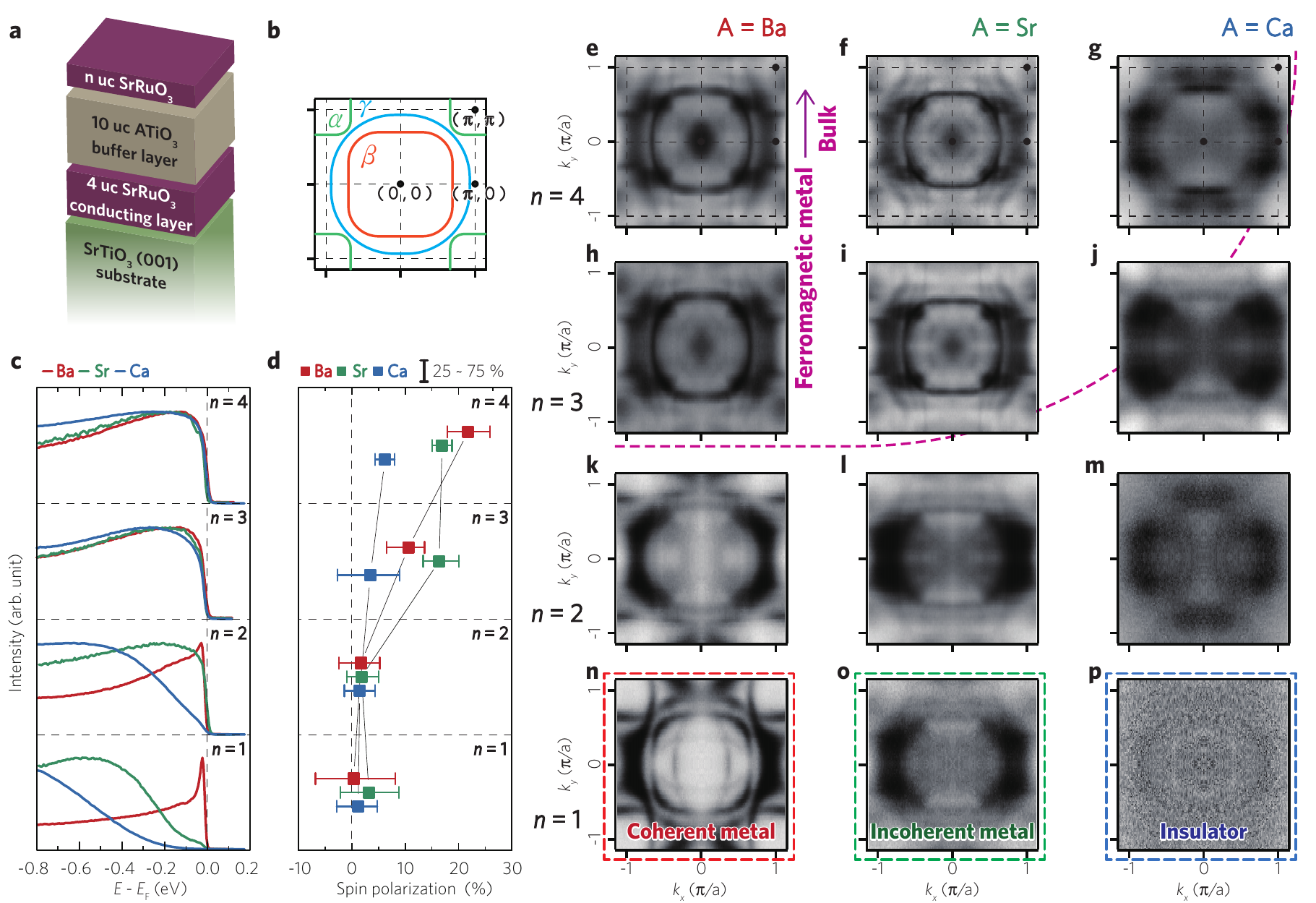}
\centering
\caption{{\bf Quantum confinement and emergence of three electronic phases in single-atomic-layer ruthenate heteroepitaxial films.} 
         {\bf a}, Schematic of charging-free ultrathin SRO heterostructures composed of a 4-UC SRO layer (conducting layer), 10-UC ATiO$_3$ (ATO, A $=$ Ba, Sr, and Ca) layer (buffer layer), and {\it n}-UC ultrathin SRO layer ({\it n} $=$ 1--4 ), sequentially grown on SrTiO$_3$ (STO) (001) substrates ({\it n}A system). 
         {\bf b}, Schematic Fermi surfaces of two-dimensional perovskite ruthenates.  
         {\bf c}, Energy distribution curves near the ($\pi$, 0) point from charging-free ultrathin SRO heterostructures obtained by angle-resolved photoemission spectroscopy (ARPES).
         {\bf d}, Spin polarization in the high-binding energy region obtained by spin-resolved ARPES measurements. The presented spin polarization is the mean value in the binding energy range of 1.0 and 1.5 eV. Error bars represent the 25--75 \% range.
         {\bf e}--{\bf p}, Symmetrized Fermi surfaces (FSs) obtained from ARPES measurements of 4Ba ({\bf e}), 4Sr ({\bf f}), 4Ca ({\bf g}), 3Ba ({\bf h}), 3Sr ({\bf i}), 3Ca ({\bf j}), 2Ba ({\bf k}), 2Sr ({\bf l}), 2Ca ({\bf m}), 1Ba ({\bf n}), 1Sr ({\bf o}), and 1Ca ({\bf p}). Ferromagnetic and non-ferromagnetic phases are separated by a purple dashed line in the background. The magnetic state of 3Ca is uncertain, considering its small spin polarization and large error bar. Coherent metal, incoherent metal, and insulator phases emerge in the 1A systems. {\bf f}, {\bf i}, {\bf l}, and {\bf o} are adapted from ref.~\cite{sohn2021observation}. 
         }
\label{fig:2}
\end{figure*}
%%%%%%%%%%%%%%%%%%%%%%%%%%%%%%%%%%%%%%%%%%%%%%%%%%%%%%%%%%%%%%

%%%%%%%%%%%%%%%%%%%%%%%%%%%%%%%%%%%%%%%%%%%%%%%%%%%%%%%%%%%%%%%
\begin{figure*}[]
\includegraphics[width=\linewidth]{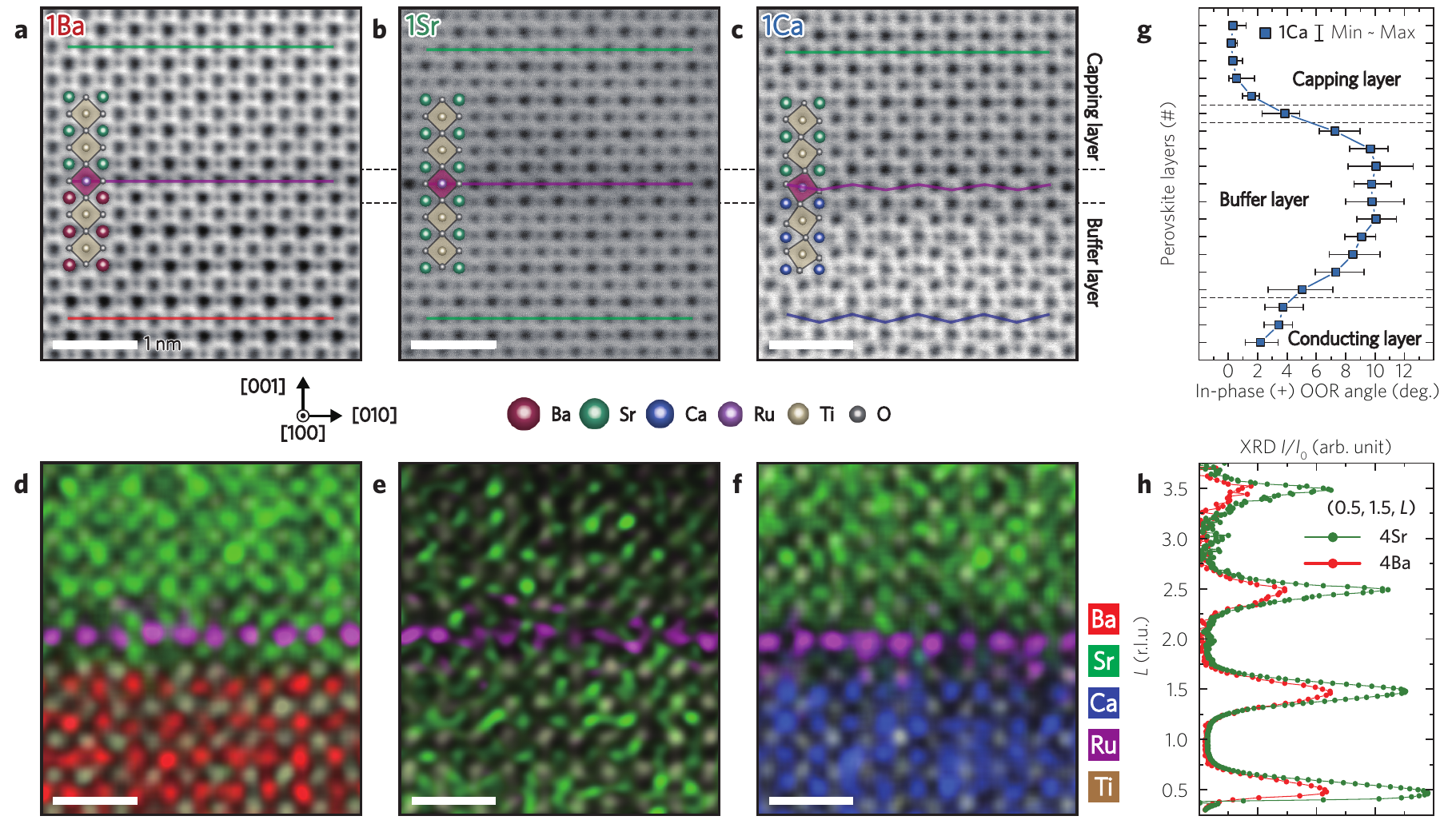}
\centering
\caption{{\bf Atomic-scale octahedral proximity effects in the single-atomic-layer ruthenates interfaced with perovskite titanates.} 
         {\bf a}--{\bf c}, Atomic-scale structural distortions near the monolayer SRO of 1Ba ({\bf a}), 1Sr ({\bf b}), and 1Ca ({\bf c}), visualized by cross-sectional annular bright-field scanning transmission electron microscopy.
         {\bf d}--{\bf f}, Atomic-scale energy-dispersive X-ray spectroscopy analysis near the monolayer SRO of 1Ba ({\bf d}), 1Sr ({\bf e}), and 1Ca ({\bf f}). Ba, Sr, Ca, Ru, and Ti are red, green, blue, purple, and beige, respectively. The zone axes are the [100] direction of the STO substrate, and the white scale bars are 1 nm long for {\bf a}--{\bf f}.
         {\bf g}, Line profile of the in-phase OOR angle obtained from 1Ca. The OOR angle values are averaged along the [010] direction and plotted along the [001] direction. Error bars represent the maximum and minimum OOR angles.
         {\bf h}, Intensities of half-integer Bragg peaks measured along the (${1 \over 2}$, ${3 \over 2}$, {\it L}) rod in the modified 4A systems (A $=$ Ba and Sr). 
         }
\label{fig:3}
\end{figure*}
%%%%%%%%%%%%%%%%%%%%%%%%%%%%%%%%%%%%%%%%%%%%%%%%%%%%%%%%%%%%%%

%%%%%%%%%%%%%%%%%%%%%%%%%%%%%%%%%%%%%%%%%%%%%%%%%%%%%%%%%%%%%%%
\begin{figure*}[]
\includegraphics[width=\linewidth]{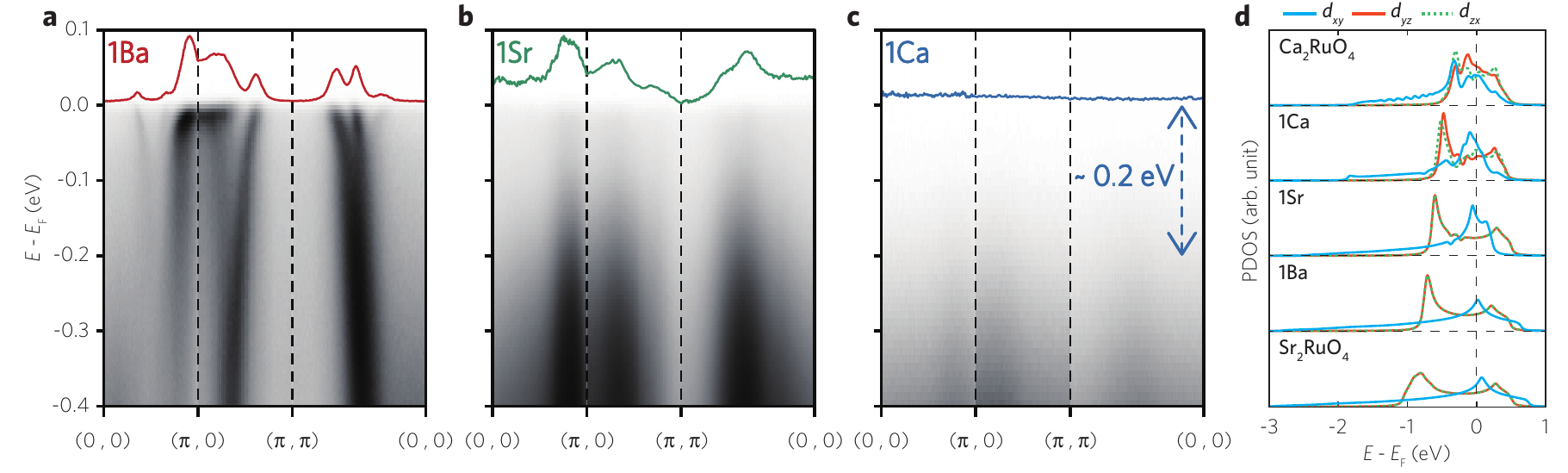}
\centering
\caption{{\bf Interfacial bandwidth control and the two-dimensional metal-to-insulator transition.} 
         {\bf a}--{\bf c}, ARPES results of energy-momentum dispersion of 1Ba ({\bf a}), 1Sr ({\bf b}), and 1Ca ({\bf c}). Momentum distribution curves at the Fermi level are drawn in red (1Ba), green (1Sr), and blue (1Ca). 
         {\bf d}, Density functional theory (DFT) calculation of the density of states (DOS) for the 1A systems. The DOS of quasi-two-dimensional ruthenates, Sr$_2$RuO$_4$ and Ca$_2$RuO$_4$, are shown for comparison.  
         }
\label{fig:4}
\end{figure*}
%%%%%%%%%%%%%%%%%%%%%%%%%%%%%%%%%%%%%%%%%%%%%%%%%%%%%%%%%%%%%%

\section*{Results}
\noindent{\bf Electronic phases of ultrathin SRO heterointerfaces}

Among the many known strongly correlated materials, ruthenates are of particular interest for their rich electronic states. In these materials, substantial Coulomb interaction ({\it U}) comparable to bandwidth ({\it W}) is present. In addition,
recent studies have emphasized the decisive role of Hund coupling ({\it J}$_H$) in their correlated electronic phases~\cite{Haule2009,Yin2011,Georges2013,lee2020interplay,HJLee2021}. In that sense, the epitaxial ruthenate heterostructures will serve as an ideal test bench for interface control of correlated electronic phases. 

The focus of this study is on the {\it n}-unit-cell ({\it n}-UC) thick SRO layers interfaced with ATiO$_3$ (ATO, A $=$ Ba, Sr, and Ca); hereinafter, we use the notation of {\it n}A. The bulk SRO is a ferromagnetic metal with a Curie temperature of around 160 K~\cite{kikugawa2015single}. The general expectation is that if {\it n} is greater than some finite number, critical thickness ({\it n}$_c$), then interface control plays no role and SRO remains a ferromagnetic metal (Fig. 1a). On the other hand, below {\it n}$_c$, it is possible to observe new phases different from the ferromagnetic metal and an interfacial crossover between them (Fig. 1b). We aim to discover these novel low-dimensional electronic phases through heteroepitaxial structural control and take the search to the single-atomic-layer limit (Fig. 1c). A more detailed description of Fig. 1 will be given as the discussion progresses.

As the A-site cation varies, ground state structures of the perovskite titanates feature ferroelectricity (A $=$ Ba)~\cite{kwei1993structures}, tetragonal distortion (A $=$ Sr)~\cite{okazaki1973lattice}, and orthorhombic distortion (A $=$ Ca, Supplementary Fig. S1)~\cite{yashima2009structural}. In the Glazer notation for oxygen octahedral rotation (OOR)~\cite{glazer1972classification}, the tetragonal and orthorhombic distortions correspond to {\it a}$^{0}${\it a}$^{0}${\it c}$^{-}$ and {\it a}$^{-}${\it b}$^{+}${\it a}$^{-}$ OOR patterns, respectively. When attached to an ultrathin SRO ({\it a}$^{-}${\it b}$^{+}${\it a}$^{-}$ OOR in bulk) layer, these characteristic ATOs exert distinct octahedral proximity effects on the film~\cite{rondinelli2012control,kim2020stabilizing}.

To explore the interfacial electronic phases, we utilized a recently established experimental method~\cite{sohn2021observation,sohn2021sign}. A key consideration is the introduction of a charging-free ultrathin oxide heterostructure schematically depicted in Fig. 2a. The heterostructures, composed of SRO ultrathin and conducting layers as well as ATO buffer layer, were fabricated via a pulsed laser deposition technique (Methods). The underlying 4-UC SRO conducting layer can supply electrons to the {\it n}-UC ultrathin layer, which prevents the charging effect during angle-resolved photoemission spectroscopy (ARPES) measurements. At the same time, the 10-UC insulating buffer layer electronically separates the conducting and ultrathin layers~\cite{sohn2021observation, yukawa2021resonant}. The key idea is that a variety of interface configurations can be achieved with the same epitaxial strain by using different buffer layers. 

We studied the electronic and magnetic properties of {\it n}A systems. Figure 2b shows a schematic Fermi surface of an ultrathin SRO film. Considering the high density of states (DOS) of the van Hov singularity (VHS) at ($\pi$,0), the electronic property of the SRO film may be partially represented by the energy distribution curves (EDCs) from ({\it k}$_x$, {\it k}$_y$) $=$ ($\pi$, 0)~\cite{sohn2021observation}. As for the magnetic properties, we employed spin-resolved ARPES. Figures 2c and 2d show EDCs and spin polarizations (averaged over the 1.5 $\leq$ {\it E}$_b$ $\leq$ 1.0 eV region, Supplementary Fig. S2), respectively, obtained from the various {\it n}A systems. For all the 4A systems, we observe considerable Fermi-level ({\it E}$_F$) spectral weight and finite spin polarization. Similar spectral weight and spin polarization have been observed for bulk and thick films~\cite{fujioka1997electronic,shai2013quasiparticle,hahn2021observation}. Specifically, the 4A systems maintain their bulk property of a ferromagnetic metal, indicating that the 4-UC thickness is excessive for interface control of the correlated electrons. Although the spin polarization of 4Ca is small compared to the other samples, the difference is not more than a quantitative change.

As the SRO layer thickness decreases, the effect of the interface is expected to gradually increase. For the 3A systems, only slight differences are observed in the electronic properties, whereas a thickness-driven ferromagnetic-to-nonmagnetic transition occurs in 3Ca. For the 2A systems, all ferromagnetism is lost due to orbital-selective quantum confinement effects~\cite{chang2009fundamental,sohn2021observation}. We begin to observe variations of the electronic properties in these nonmagnetic phases, where coherency of the quasiparticles are slightly enhanced (2Ba), slightly suppressed (2Sr), and strongly suppressed (2Ca). Finally, the interfacial effect is maximized in the single-atomic-layer limit of the 1A systems, resulting in a two-dimensional (2D) metal-to-insulator transition. The quasiparticle of 1Ba is well-defined and comparable to the 
representative Fermi liquid Sr$_2$RuO$_4$~\cite{damascelli2000fermi}. Strong incoherency is observed in 1Sr, presumably due to quantum confinement effects and the VHS~\cite{sohn2021observation}. The spectra from 1Ca show a finite insulating gap. 

These thickness- and interface-dependent electronic properties are better revealed and summarized in the Fermi surface (FS) maps presented in Figs. 2e--p. We label the FSs of ultrathin SRO films as $\alpha$, $\beta$, and $\gamma$ bands, as shown in Fig. 2b. $\alpha$ and $\beta$ bands, as well as somewhat broad $\gamma$ bands, are seen in thick samples (4A and 3A). As the 2D limit is approached, the FSs of {\it n}Ba systems show a similar behavior to the archetypal 3 FS bands in Sr$_2$RuO$_4$ (Fig. 2b). In contrast, the FSs of {\it n}Sr and {\it n}Ca systems become gradually broad. Especially, 1Ca does not show any spectral weight at {\it E}$_F$.\\

\noindent{\bf Structure characterization of the heterostructures}

We carried out a series of structural analyses, focusing on the 1A systems (Fig. 1c). We first examine the surface symmetry of the 1A systems using low-energy electron diffraction (LEED, Supplementary Fig. S3) which was performed {\it in situ} in the ARPES chamber. The LEED patterns of the 1A systems show $\sqrt{2}\times\sqrt{2}$ (1Ba and 1Sr) and 2$\times$2 (1Ca) surface reconstructions, which are assigned to the {\it a}$^{0}${\it a}$^{0}${\it c}$^{-}$ and {\it a}$^{-}${\it b}$^{+}${\it c}$^{-}$ OOR patterns, respectively (Methods).

Cross-sectional scanning transmission electron microscopy (STEM) provides direct atomic-scale visualization of the octahedral arrangements of the 1A systems along the zone axis of the in-plane [100] direction. The 1A systems for STEM measurement were capped with 10-UC SrTiO$_3$ (STO) layers (capping layer) to protect the topmost monolayer SRO. All measurements were conducted at room temperature {\it ex situ}. Figures 3a--c show annular bright-field STEM (ABF-STEM) images near the monolayer SRO of the 1A systems. The high sensitivity of the ABF-STEM for light elements (i.e., oxygen) enables fine resolution of the oxygen octahedral structures. We confirm the chemical composition and associated interface structures using atomic-scale energy-dispersive X-ray spectroscopy; the results are displayed in Figs. 3d--f. The sharp interfaces are further supported by the STEM line profiles shown in Supplementary Fig. S4.  

The STEM pictures show that the OORs of all the 1A systems are similar to those of the corresponding ATO buffer layer, experimentally confirming the structural proximity effects. Figures 3a,b show that, in 1Ba and 1Sr, all B-site cations (i.e., Ti and Ru) and oxygen ions form straight lines toward the [010] direction. It indicates that these two heterostructures do not exhibit in-plane-oriented OOR, similar to the cases of bulk BaTiO$_3$ (BTO) and STO. This observation is consistent with the LEED analysis of {\it a}$^{0}${\it a}$^{0}${\it c}$^{-}$ that exhibits only out-of-plane-oriented OOR. In the CaTiO$_3$ (CTO) layer of 1Ca, oxygen ions are significantly displaced from the central position between two adjacent Ti ions. The Ti and O ions here form a zigzag-like pattern along the [010] direction. This pattern propagates, through the monolayer SRO, up to the first two-UC of the STO capping layer (Fig. 3g and Supplementary Fig. S5). We attribute the zigzag-like pattern to the in-phase ($+$) OOR of the {\it a}$^{-}${\it b}$^{+}${\it c}$^{-}$. Although the STEM measurements were conducted at room temperature, we assume that 1Ba and 1Ca exhibit similar characteristics at low temperatures as BTO and CTO do not have any change in the OOR at low temperatures~\cite{kwei1993structures,yashima2009structural}.

Although the structural characterizations thus far cannot distinguish 1Ba and 1Sr, half-integer Bragg peak X-ray diffraction (XRD) analysis implies that these two samples have distinct OOR angles. For XRD analysis, we used modified 4A systems, and the measurements were conducted at room temperature {\it ex situ} (Methods). Figure 3h shows the intensities of half-integer Bragg peaks along the (${1 \over 2}$, ${3 \over 2}$, {\it L}) rod of the modified 4Ba and 4Sr in pseudo-cubic (pc) notation [(${1 \over 2}$, ${5 \over 2}$, {\it L}) and (${3 \over 2}$, ${5 \over 2}$, {\it L}) rods are shown in Supplementary Figs. S6--S8]. These Bragg peaks are known to reflect the {\it a}$^{0}${\it a}$^{0}${\it c}$^{-}$ OOR~\cite{glazer1975simple}. The modified 4Ba shows smaller normalized intensities than the modified 4Sr. We qualitatively interpret this as a suppressed {\it a}$^{0}${\it a}$^{0}${\it c}$^{-}$ OOR. Because the interface structural coupling is the maximum in the single-atomic-layer limit, we expect an even larger OOR angle difference between 1Ba and 1Sr.

To summarize the structural analysis of the 1A systems, octahedral proximity effects induce the monolayer SRO to adopt a small angle {\it a}$^{0}${\it a}$^{0}${\it c}$^{-}$ OOR (1Ba), larger angle {\it a}$^{0}${\it a}$^{0}${\it c}$^{-}$ OOR (1Sr), and {\it a}$^{-}${\it b}$^{+}${\it c}$^{-}$ OOR (1Ca), as schematically depicted in Fig. 1c. Previously, it was attempted to show how the OOR affects the electronic structure through dimension (thickness) and strain dependence studies~\cite{schutz2017dimensionality,dymkowski2014strain}. However, the dimensionality and strain themselves have a significant effect on the electronic structure~\cite{sohn2021observation,burganov2016strain,sunko2019direct}. All heterostructures considered here are fully epitaxially strained to the STO (001) substrate (Supplementary Fig. S9), and the 2D electronic transition occurs exclusively at the {\it n} $=$ 1 thickness. Our experimental design realizes an appropriate separation of variables for OOR, dimensionality, and strain. The unique interfacial configuration is the sole source of the electronic transition in the monolayer SRO.\\

\noindent{\bf 2D metal-to-insulator transition in the monolayer SRO}

Small differences in the oxygen octahedral arrangements contribute to significant changes in the 2D electronic structure. Figures 4a--c display the energy--momentum ({\it E}--{\it k}) dispersion of the 1A systems. Consistent with the sharp FS (Fig. 2n), 1Ba shows a coherent {\it E}--{\it k} dispersion near {\it E}$_F$. In particular, the clear VHS at the ($\pi$, 0) point highlights the DOS singularity and 2D character. 1Sr and 1Ca consistently show broad spectra and incoherent dispersion. Spectral weight of 1Ca appears only below {\it E}$_b$ $=$ 0.2 eV. Our experimental results indicate that the interface engineering yields coherent metal (1Ba), incoherent metal (1Sr), and insulator (1Ca) phases from the bulk ferromagnetic metal. The strong incoherency in 1Sr and 1Ca implies the presence of a strong electronic correlation~\cite{damascelli2003angle}.

To understand the strongly correlated electronic phases of the 2D ruthenates, we employed a combined approach of density functional theory (DFT) and dynamical mean-field theory (DMFT). First, we performed DFT calculations to simulate the heterostructures of our samples (Methods and Supplementary Fig. S10). Consistent with our structural analysis, DFT calculations reproduce the observed OORs, showing {\it a}$^{0}${\it a}$^{0}${\it c}$^{-}$ (1Ba and 1Sr) and {\it a}$^{-}${\it b}$^{+}${\it c}$^{-}$ OORs (1Ca). Here, we notice that the angle of the out-of-plane-oriented rotation was significantly suppressed for 1Ba, in comparison to 1Sr. Unlike the ground state of the STO ({\it a}$^{0}${\it a}$^{0}${\it c}$^{-}$), the bulk BTO does not exhibit any OOR over all temperature ranges~\cite{kwei1993structures,okazaki1973lattice}. This difference is expected to suppress the OOR of 1Ba at the heterointerfaces. The OOR distortion involves the bent metal-oxygen-metal bonding of neighboring sites, which is known to suppress electron hopping between metal ions~\cite{guzman2019cooperative}. Even in the 1A systems, suppression or enhancement of the OOR systematically control {\it W} of 2D ruthenates. Figure 4d shows the DFT orbital-projected DOS without consideration of Hubbard or Hund interaction. We can observe the effect of the bandwidth control in our thin film samples alongside other quasi-2D ruthenates (Sr$_2$RuO$_4$ and Ca$_2$RuO$_4$, Supplementary Table 1). 

Ruthenates are known for their subtle balance between {\it U}, {\it J}$_H$, and the {\it W}~\cite{Haule2009,Yin2011,Georges2013,lee2020interplay,HJLee2021}. Here, the {\it U} and {\it J}$_H$ are almost intrinsic constants for the Ru atom, and the change in the {\it W} effectively enhances or suppresses the effect of the correlation (Hubbard or Hund interaction). Through octahedral proximity and corresponding bandwidth control, we expect the electronic phases of 2D ruthenates to progress through that of a coherent Fermi liquid, an incoherent Hund metal, and an antiferromagnetic Mott insulator as depicted in the phase diagram in Fig. 1b. The phase diagram was schematically drawn with reference to previous studies~\cite{Georges2013,dang2015electronic}.\\

%%%%%%%%%%%%%%%%%%%%%%%%%%%%%%%%%%%%%%%%%%%%%%%%%%%%%%%%%%%%%%%
\begin{figure*}[]
\includegraphics[width=\linewidth]{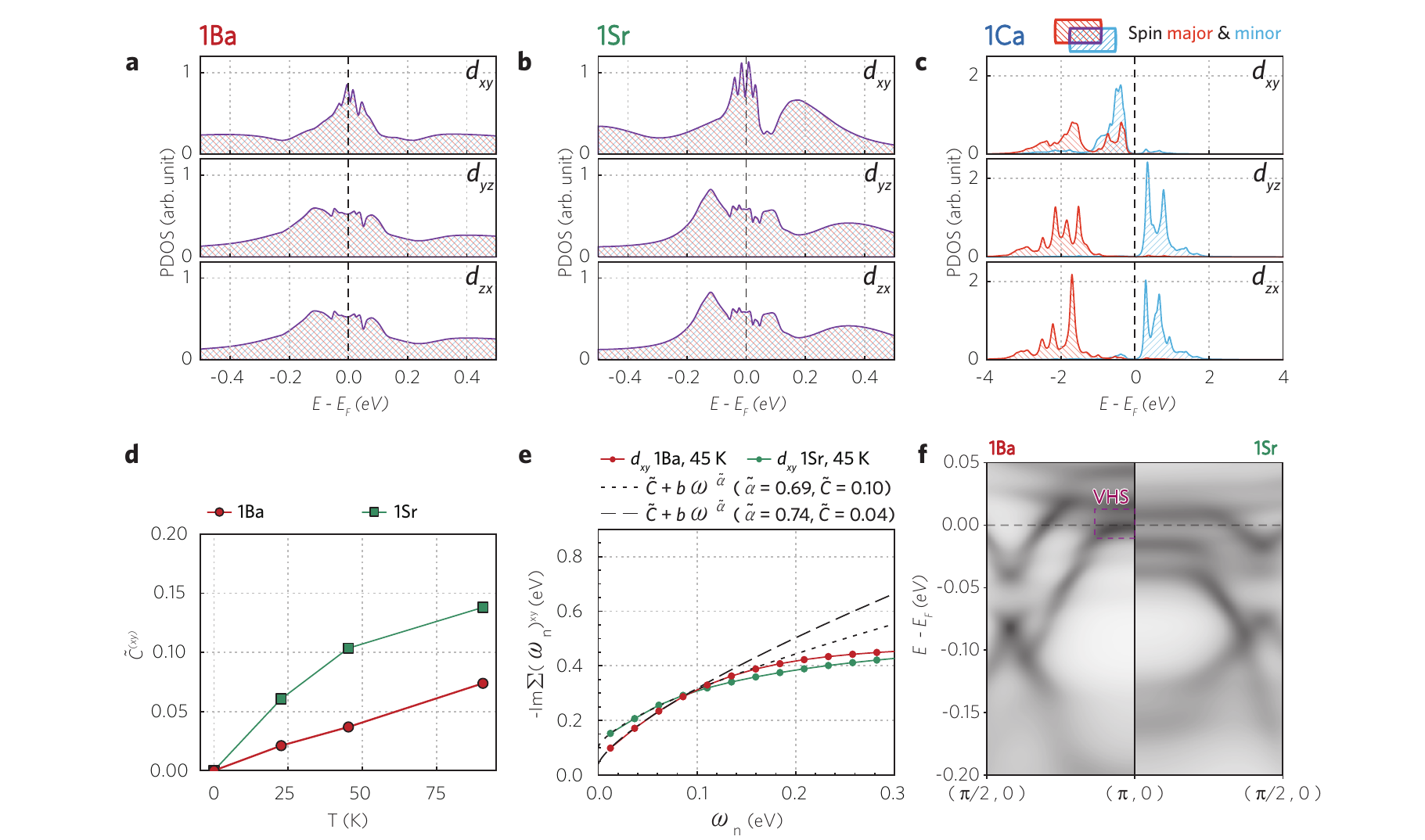}
\centering
\caption{{\bf Interfacial crossover between Fermi liquid, Hund metal, and Mott insulator phases.} 
         {\bf a}--{\bf c}, Orbital-projected DOS from dynamical mean-field theory (DMFT) calculations of 1Ba ({\bf a}), 1Sr ({\bf b}), and 1Ca ({\bf c}). The red (blue) striped area corresponds to the DOS of spin majority (minority).
         {\bf d}, Scattering rate $\tilde{C}$ of the {\it d}$_{xy}$ component of 1Ba and 1Sr as a function of temperature.
         {\bf e}, Matsubara self-energies of $d_{xy}$  at 45 K with fitting lines of $A\omega^\alpha+\tilde{C}$. 
         {\bf f}, Momentum-resolved DMFT spectral functions of 1Ba and 1Sr.}
\label{fig:5}
\end{figure*}
%%%%%%%%%%%%%%%%%%%%%%%%%%%%%%%%%%%%%%%%%%%%%%%%%%%%%%%%%%%%%%

\noindent{\bf Crossover between Fermi liquid, Hund metal, and Mott insulator phases}

We attempt to verify the correlated phases using DMFT calculations. Throughout the calculation, we assumed {\it U} $=$ 2.3 eV and {\it J}$_H$ $=$ 0.4 eV, which are known to yield a good agreement between theoretical and experimental studies~\cite{Mravlje2011,Tamai2019}. Figures 5a--c show the DMFT orbital-projected DOS of the 1A systems.
1Ca with a small {\it W} ($=$ 2.4 eV) shows no DOS at $E_F$. The $d_{xy}$ orbital is fully occupied, while the {\it d}$_{yz}$ and {\it d}$_{zx}$ orbitals exhibit an open Mott gap. This type of band filling is identical to the filling in Ca$_2$RuO$_4$ with the {\it a}$^{-}${\it a}$^{-}${\it c}$^{+}$ OOR.

For 1Sr with a slightly larger {\it W} ($=$ 2.8 eV), the effective correlation is insufficient to open the Mott gap but significant. In this context, multi-orbital systems have been known to exhibit strong correlations associated with the {\it J}$_H$. The {\it J}$_H$-driven correlations lead to crossover from a Fermi liquid to an incoherent metallic phase (i.e., Hund metal) over a wide range of temperatures. In this incoherent metallic phase, the spin is not fully screened due to the strong Hund interaction and acts as a scattering center, resulting in incoherent quasiparticles~\cite{Georges2013}.

To understand the coherence--incoherence crossover observed in our 1Ba and 1Sr, we analyzed the low-energy behavior of the imaginary part of the Matsubara axis self-energy. We fitted the self-energy to the form $-\mathrm{Im}\Sigma_{\mu\mu} (i\omega) \sim A\omega^{\tilde{\alpha}^{(\mu)}} + \tilde{C}^{(\mu)}$, where $\mu \in t_{2g}$~\cite{Werner2008,Werner2012}.
In an ideal Fermi liquid, $\mathrm{Im}\Sigma_{\mu\mu} (i\omega)$ exhibits a linear regime at low energy ($\tilde{\alpha}=1$, $\tilde{C}=0$). However, if the {\it J}$_H$ is sufficiently large and the temperature is sufficiently high, the spin is not fully screened
and reaches the region where the frozen moment occurs. Then, the intercept $\tilde{C}$ (i.e., the scattering rate) becomes nonzero~\cite{Werner2008,Werner2012}. This is a typical self-energy behavior evident in a Hund metal phase. In our calculations, both 1Ba and 1Sr show non-zero intercepts $\tilde{C}$ in a finite temperature regime (Fig. 5d). Nevertheless, the correlation effects are greatly weakened by the large {\it W} of 1Ba. As a result, in the very low temperature range ($\sim$10 K) where our ARPES measurements were performed, 1Ba remains very close to a Fermi liquid ($\tilde{C}$ is almost 0) while 1Sr exhibits comparatively stronger incoherence (Figs. 5d,e). These features are visible in both DOS (Figs. 5a,b) and momentum-resolved spectral functions (Fig. 5f) as well as FS calculations (Supplementary Fig. S11).

It should be noted that the incoherent metal phase, observed particularly in 1Sr, cannot be found over the entire temperature range without considering the {\it J}$_H$; the system remains as an ideal Fermi liquid (Supplementary Fig. S12). Additionally, from the perspective of Hund metal, this self-energy behavior shows a very strong orbital dependence. The non-zero intercept shown in Fig. 5d is only evident in the {\it d}$_{xy}$ component of self-energy, while the {\it d}$_{yz}$ and {\it d}$_{zx}$ components behave as Fermi liquids in all areas. Therefore, we conclude that these three types of correlated phases are Fermi liquid (1Ba), Hund metal (1Sr), and Mott insulator (1Ca).

\section*{Discussion}

It is important to revisit the characteristics of 1Ba (Fermi liquid) that distinguish it from its adjacent phases, 1Sr (Hund metal) and thick SRO films (ferromagnetic metal). As discussed earlier, Hund metallicity appears as a coherence--incoherence crossover phase. In particular, the presence of the VHS near {\it E}$_F$ effectively enhances the {\it J}$_H$ and incoherency~\cite{lee2020interplay,karp2020sr}. The bare bands of both 1Ba and 1Sr exhibit a VHS near {\it E}$_F$ ("M" $=$ ($\pi$, 0) point in Supplementary Fig. S13)~\cite{sohn2021observation}. Both our theory (Fig. 5f) and experiment (Figs. 4a,b, and Supplementary Fig. S14) consistently indicate that only 1Ba maintains its coherent VHS near {\it E}$_F$. While the momentum-resolved spectral function of 1Sr becomes incoherent, especially near the VHS (Fig. 5f), the incoherency in the experimental spectra is considerable within a rather extended {\it E}--{\it k} space (Fig. 4b). Note that there still is room to improve, including the precise experimental values for OOR angle and electron number as well as the exact values for the interaction parameters ({\it U} and {\it J}$_H$). Further studies on more refined structural analysis and theoretical modeling are highly desired to obtain better quantitative agreements.

A recent study demonstrated that thick SRO films in the ferromagnetic metal phase have coherent dispersions at low temperatures~\cite{hahn2021observation}. With spin-dependent electronic correlations, the films show substantial coherence--incoherence crossover as the temperature rises. In contrast, 1Ba without ferromagnetism maintains its coherent dispersion up to 100 K (Supplementary Fig. S14). Therefore, 1Ba is more analogous to an archetypal Fermi liquid, Sr$_2$RuO$_4$~\cite{shen2007evolution}, rather than the ferromagnetic metal phase. Even at high temperatures, 1Ba shows the coherent VHS near {\it E}$_F$.\\

Lastly, we discuss the necessity of using a single-atomic-layer in the interface control of the SRO heterostructures. The excessive thickness of the SRO layer weakens the desired interface control from three perspectives (Fig. 1a). First, the bulk ferromagnetic metal phase becomes dominant~\cite{dang2015electronic}. In our study, ultrathin SRO films are electronically relaxed to the bulk ferromagnetic phase only after 2- or 3-UC thickness. Second, the three 2D correlated phases we obtained are closely related to the quantum confinement effect. The quantum confinement on ultrathin SRO layers is selective for {\it d}$_{yz}$ and {\it d}$_{zx}$ orbitals. The 1A systems feature coherent VHS of {\it d}$_{xy}$ band (1Ba), orbital-dependent Hund metallicity (1Sr), and coexistence of band- and Mott-insulating orbitals (1Ca). Orbital-selective quantum confinement could facilitate the stabilization of 2D correlated phases and bandwidth control over them. Third, the octahedral proximity effect is also relaxed after several layers. Although our heterostructures do not show a thickness-dependent transition of the OOR pattern in the {\it n}A systems, the control over the OOR angle may have been relaxed (Fig. 3g). The length scale of the OOR control is typically only several atomic layers~\cite{kan2016tuning,kim2020stabilizing,dominguez2020length}. Our interpretation is partially consistent with experimental observations. The interfacial crossover of correlated phases is exclusively possible through the combined engineering of orbital-selective quantum confinement and giant octahedral proximity effects.\\

\noindent{\bf Outlook}

Correlated oxide interfaces produce a variety of electronic phases, but the length scale of the control is often limited. Our ARPES study of charging-free ultrathin SRO heterostructures suggests a method for investigating the electronic phases at ultrathin oxide interfaces. The interface control in our study is only effective in the single-atomic-layer limit, but the local modification is unexpectedly large. When approaching the single-atomic-layer limit, interface engineering can induce transitions from a bulk-like ferromagnetic metal to Fermi liquid, Hund metal, and Mott insulator phases. The cooperation of orbital-selective quantum confinement and octahedral proximity effects have implemented 2D bandwidth control of strongly correlated electron systems. Although we only considered structural control, the application of numerous buffer layers with distinct properties (e.g., magnetic buffer layer) is the focus of follow-up studies. We also would like to point out that our work on single-atomic-layer films is on par with studies on exfoliated monolayers of van der Waals materials but with different perspectives. Therefore, atomically thin oxide heterointerfaces are expected to reveal emergent electronic phases that have not yet been explored and their studies can be complementary to current research on 2D materials.

\section*{Methods}
\noindent\textbf{Fabrication and characterization of the oxide heterostructures}

Oxide heterostructures composed of SrRuO$_3$ (SRO), BaTiO$_3$ (BTO), SrTiO$_3$ (STO), and CaTiO$_3$ (CTO) were epitaxially grown on (001)-oriented STO single crystal substrates (Shinkosha Co., Ltd.) via pulsed laser deposition (Pascal Co., Ltd.). Before growth, the STO substrate was dipped in deionized water and sonicated for 30 min. The substrate was subsequently annealed in the growth chamber {\it in situ}; the annealing temperature, background oxygen partial pressure, and annealing time were $1,050\,^{\circ}{\rm C}$, 5.0~$\times$10$^{-6}$~Torr, and 30 min, respectively. Polycrystalline SRO, BaTi$_{1.2}$O$_{3.4}$, SrTiO$_3$, and Ca$_{1.1}$TiO$_{3.1}$ targets (Toshima manufacturing Co., Ltd.) were ablated using a KrF excimer laser (Coherent Inc.). For the growth of SRO films, the substrate temperature, background oxygen partial pressure, and laser energy density were kept at $670\,^{\circ}{\rm C}$, 100~mTorr, and 1.9~J/cm$^{2}$, respectively. For the growth of BTO, STO, and CTO films, the substrate temperature and background oxygen partial pressure were kept at $670\,^{\circ}{\rm C}$ and 10~mTorr, respectively.

The cation stoichiometry of perovskite titanate thin films is highly sensitive to the laser energy density, which directly influences the quality of multi-layer oxide heterostructures. We optimized the laser energy density for BTO, STO, and CTO by growing 20 nm-thick films on lattice-matched GdScO$_3$ (110) [GSO (110)], STO (001), and (LaAlO$_3$)$_{0.3}$--(Sr$_2$AlTaO$_6$)$_{0.7}$ (001) [LSAT (001)] substrates (Crystec GmbH), respectively. Supplementary Figures S1d--f show the X-ray diffraction (XRD) results of the BTO, STO, and CTO thin films with varying laser energy density. The BTO film on GSO (110) shows clear Kiessig fringes only within a finite range of the laser energy density. The poor crystallinity outside this range is related to cation off-stoichiometry~\cite{matsubara2014single}. The XRD peak position of our optimal BTO thin film grown at 1.0~J/cm$^{2}$  
indicates a slightly larger out-of-plane lattice constant, compared to earlier molecular beam epitaxy studies~\cite{matsubara2014single,choi2004enhancement}. The off-stoichiometry of the STO films is known to cause lattice expansion~\cite{ohnishi2008defects,brooks2009growth}. We optimized the STO films by minimizing the lattice expansion. The growth optimization of CTO films based on the surface morphology and reflection high-energy electron diffraction is reported elsewhere~\cite{kim2020stabilizing}. When the laser energy density was varied, the optimal CTO film strained on the LSAT (001) substrate showed the minimum out-of-plane lattice constant~\cite{roth2018temperature}.

A high-resolution X-ray diffractometer (AXS D8 with a Vantec line-detector, Bruker, USA) was used for general structural characterization including line scan and reciprocal space mapping. The epitaxial state of the 1A systems was examined (Supplementary Fig. S9). The XRD peak intensity of the SRO layer is weak, due to the small number of layers: a 4-unit-cell (4-UC) conducting layer and monolayer SRO. Nevertheless, the XRD peaks of the 10-UC buffer layers (BTO and CTO) are observable, and the {\it H} position is identical to the adjacent substrate peak. Because the ATO buffer layers are above the 4-UC SRO, we conclude that all heterostructures are fully strained on the STO (001) substrate.

Low-energy electron diffraction (LEED) was conducted simultaneously with the angle-resolved photoemission spectroscopy (ARPES) {\it in situ} (Supplementary Fig. S3). Among the known oxygen octahedral rotation (OOR) patterns of the constituting perovskite materials, the $\sqrt{2}\times\sqrt{2}$ reconstruction corresponds to the {\it a}$^{0}${\it a}$^{0}${\it c}$^{-}$ pattern of STO ({\it T} $\leq$ 106 K), CTO (1,512 $\leq$ {\it T} $\leq$ 1,636 K), and SRO (820 $\leq$ {\it T} $\leq$ 950 K), and the 2$\times$2 reconstruction corresponds to the {\it a}$^{-}${\it b}$^{+}${\it a}$^{-}$ pattern of CTO ({\it T} $\leq$ 1,512 K) and SRO ({\it T} $\leq$ 820 K)~\cite{okazaki1973lattice,yashima2009structural,kennedy1998high}. Because no other low-symmetric OOR patterns have been reported from bulk ATOs and SRO, we did not consider other possibilities. The LEED patterns of the {\it n}A surfaces did not exhibit thickness dependence, at least up to 4-UC thickness (Supplementary Fig. S3). Note that the orthorhombic {\it a}$^{-}${\it b}$^{+}${\it a}$^{-}$ becomes monoclinic {\it a}$^{-}${\it b}$^{+}${\it c}$^{-}$ under (001)-oriented epitaxial strain~\cite{vailionis2008room}.

Schematic figures of atomic structures were drawn using VESTA~\cite{momma2011vesta}.\\

\noindent\textbf{\textit{In-situ} ARPES}

{\it In-situ} ARPES measurements were performed using a home lab system equipped with a Scienta DA30 analyzer and a discharge lamp from Fermi instrument. He-I$\alpha$ ($hv=21.2$~eV) light was used. All ARPES data were measured at 10~K, with the exception of the high-temperature ARPES data (Supplementary Fig. S14) measured at 100~K. LEED patterns were recorded after ARPES measurements at 10~K (Supplementary Fig. S3). 

Spin polarization was measured with spin-resolved ARPES using a home-lab ARPES system (Fig. 2d and Supplementary Fig. S2). The system was equipped with a SPECS PHOIBOS 225 analyzer and a very low energy electron diffraction spin detector. For the spin detector, an oxidized iron film deposited on W(100) was used as the scattering target. He-I$\alpha$ ($hv=21.2$~eV) light was used as the light source. Thin films were field-cooled from 80 to 10~K using a permanent magnet. To clean the surfaces of SRO thin films, we post-annealed them at $550\,^{\circ}{\rm C}$ for 10~min~\cite{sohn2021observation}.\\

\noindent\textbf{DFT$+$DMFT calculations}

We carried out DFT calculations within the Perdew--Burke--Ernzerhof exchange-correlation functional revised for solids using VASP code~\cite{Kresse1996,Kresse1999}.
To describe the experimental situation, we prepared three slab geometries --- (i) 1Ba: 1SRO/5BTO/3STO/5BTO/1SRO, (ii) 1Sr: 1SRO/4STO/1SRO, and (iii) 1Ca: 1SRO/4CTO/1SRO with 20 \AA~ of vacuum. These geometries are symmetrical with respect to the middle layer as shown in Supplementary Fig. S10.
We used a 600 eV plane wave cut-off energy and 6$\times$6$\times$1 k-points for all DFT calculations.
The in-plane lattice constant was fixed at the experimental value of STO (3.905\AA).
The internal atomic positions of the three layers closest to the surface were fully relaxed until the maximum force was below 5 meV\AA$^{-1}$,
while the remaining bulk part ions were constrained during relaxation to ensure calculation stability.
Maximally localized Wannier functions~\cite{MLWF} were obtained from the DFT $t_{2g}$ bands, as shown in Supplementary Fig. S13.
We performed single-site DMFT calculations on top of the Wannier Hamiltonian using an exact diagonalization solver.
We used a rotationally invariant Slater Kanamori interaction for t$_{2g}$ impurity orbitals, with 18 bath levels in the impurity Hamiltonian. To compare Hund metallic character of 1Ba and 1Sr, we used electron filling {\it n} $=$ 3.85, which reproduces the position of the VHS clearly visible in the ARPES spectrum of 1Ba.\\

\noindent\textbf{STEM measurement}

Cross-sectional STEM specimens were prepared via Ga ion milling in a focused ion beam with an FEI Helios 650 FIB and further thinned by focused Ar ion milling using a Fischione NanoMill 1040. STEM images and EDS results were acquired using a Thermo Fisher Scientific Themis Z aberration-corrected microscope with a 300 kV primary beam energy and 25.1 mrad semi-convergence angle.

Annular bright-field (ABF) and high-angle annular dark-field (HAADF) STEM images, including the entire STO-capped charging-free monolayer SRO heterostructures, are shown in Supplementary Fig. S4. The ABF STEM is suitable for imaging oxygen ions, and the HAADF STEM more clearly shows the contrast between layers. The HAADF STEM intensity plots across the heterostructures (dashed box) corroborate the sharp epitaxial interface structures, consistent with EDS mapping results discussed in the main text.     \\

\noindent\textbf{Crystal truncation rod measurements}

Crystal truncation rods (CTRs) were measured with a surface XRD technique using a six-circle diffractometer under an incident photon energy of 15.5~keV at beamline 33-ID-D of the Advanced Photon Source, Argonne National Laboratory. The X-ray beam was focused by a pair of Kirkpatrick--Baez mirrors down to a size of 50~$\mu m$ (vertical)~$\times$~100~$\mu m$ (horizontal). The resulting two-dimensional diffraction images in reciprocal space were collected using a Dectris PILATUS 100 K area detector. Experimental integer and half-integer CTR intensities were extracted from detector images after proper background subtraction, normalization, and geometric/polarization corrections. During room temperature measurements, samples were kept under dry helium gas flow in a sealed Kapton dome.

To investigate and compare OOR patterns of ultrathin SRO layers on BTO and STO buffer layers, half-integer Bragg peaks were measured. For high-resolution structural analysis, we used modified 4Ba (4Sr) composed of 2-UC SRO conducting layers, 6-UC BTO (STO) buffer layers, 4-UC SRO ultrathin layers, and STO (001) substrates. We found that only the $1/2(hkl)$ peaks (where $h$, $k$, and $l$ are odd integers and $h$~$\neq$~$k$) appeared in both heterostructures (Supplementary Figs. S6--8). Thus, both heterostructures have the same OOR patterns corresponding to the $a^0a^0c^-$ (i.e., a I4/$mcm$ space group)~\cite{glazer1975simple}.
	
By comparing the intensities of half-Bragg peaks between two heterostructures, the magnitude of OOR was estimated~\cite{qiao2015dimensionality, fister2014octahedral}. As shown in Fig. 3h and Supplementary Figs. S6--S8, the intensities of $1/2(hkl)$ peaks (where $h$, $k$, and $l$ are odd integers and $h$~$\neq$~$k$) were weaker in the modified 4Ba than in the modified 4Sr, indicating that the OOR angles were generally smaller in the ultrathin SRO layers on the BTO buffer layers.

The atomic structures of the above-described heterostructures were determined using a three-dimensional electron density map reconstructed from a large set of CTRs. An iterative phase retrieval technique, known as coherent Bragg rod analysis (COBRA)~\cite{yacoby2002direct, kumah2014tuning, disa2020high}, was used to derive a real-space electron density map. COBRA-derived electron density maps for the modified 4Sr and 4Ba, as well as the electron density profiles and layer-dependent lattice parameters, are shown in Supplementary Figs. S15 and S16. These findings highlight the structural integrities of the deposited films and validate our analysis of OOR using half-order CTRs.\\

\section*{Acknowledgements}
We gratefully acknowledge insightful discussions with Kookrin Char, Kee Hoon Kim, Hongki Min, Bohm Jung Yang, Sang Mo Yang, Seo Hyoung Chang, Changhee Sohn, Yeong Jae Shin, Wenzheng Wei, Kidae Shin, and Joseph Falson. This work is supported by the Institute for Basic science in Korea (Grant No. IBS-R009-D1, IBS-R009-G2, IBS-R024-D1). We acknowledge the support from the Korean government through National Research Foundation (2017R1A2B3011629). Cs-corrected STEM works were supported by the Research Institute of Advanced Materials (RIAM) in Seoul National University. Work at Yale was supported by the Air Force Office of Scientific Research (AFOSR) under Grant No. FA9550-21-1-0173. Surface X-ray diffraction experiments were performed at the beam line 33-ID-D of the Advanced Photon Source, a U.S. Department of Energy (DOE) Office of Science User Facility operated for the DOE Office of Science by Argonne National Laboratory under Contract No. DE-AC02-06CH11357. AG was supported by the National Research Foundation of Korea (NRF) under 2021R1C1C1010429. \\

%\bibliographystyle{naturemag}
%\bibliography{ref}

\clearpage

%%%%%%%%%%%%%%%%%%%
{
\renewcommand{\thetable}{1}
\begin{table*}
\centering
\caption{
	\label{table:comparison} 
	Bandwidths ($W_{xy}$ and $W_{yz/zx}$) estimated from Wannier Hamiltonian.
      }
        \begin{tabular}{ccD{.}{.}{2.3}D{.}{.}{2.3}}\hline\hline
          Systems &  \multicolumn{1}{c}{$W_{xy}$ (eV)} & \multicolumn{1}{c}{~~~~~$W_{yz/zx}$ (eV)} \\\hline
          ~~~~Ca$_2$RuO$_4$~~~~ & 2.3 & 1.1\\
          1Ca & 2.4 & 1.3\\
          1Sr & 2.8 & 1.4\\
          1Ba & 3.5 & 1.5\\
          Sr$_2$RuO$_4$ & 3.8 & 1.8\\
    \hline\hline
  \end{tabular}
\end{table*}
}
%%%%%%%%%%%%%%%%%%%%%%%%

\renewcommand{\thefigure}{S1}
\begin{figure*}[h!]
\centering
\includegraphics[width=\linewidth]{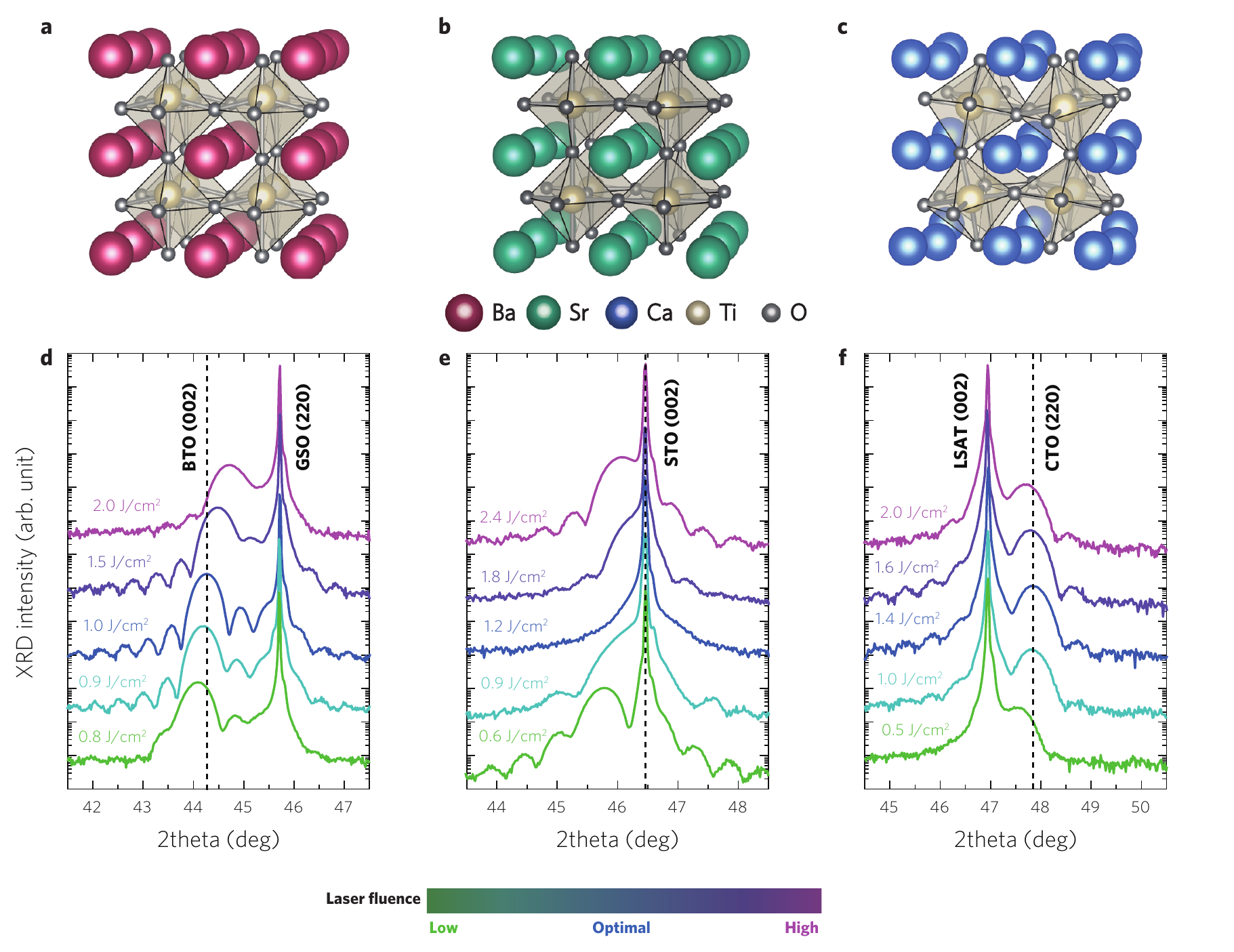}
\caption{{\bf Growth optimization of perovskite titanates thin films.} {\bf a}--{\bf c}, Crystal structures of BaTiO$_3$ (BTO, {\bf a}), SrTiO$_3$ (STO, {\bf b}), and CaTiO$_3$ (CTO, {\bf c}), which have ferroelectric, tetragonal, and orthorhombic distortions, respectively. {\bf d}--{\bf f}, X-ray diffraction (XRD) 2theta--theta scan of BTO, STO, and CTO thin films grown on GdScO$_3$ (110) [GSO (110)], STO (001), and (LaAlO$_3$)$_{0.3}$--(Sr$_2$AlTaO$_6$)$_{0.7}$ (001) [LSAT (001)] substrates, respectively. Optimal thin films were achieved by controlling the laser energy density for pulsed laser deposition.}
\label{Ch1}
\end{figure*}

\renewcommand{\thefigure}{S2}
\begin{figure*}[h!]
\centering
\includegraphics[width=\linewidth]{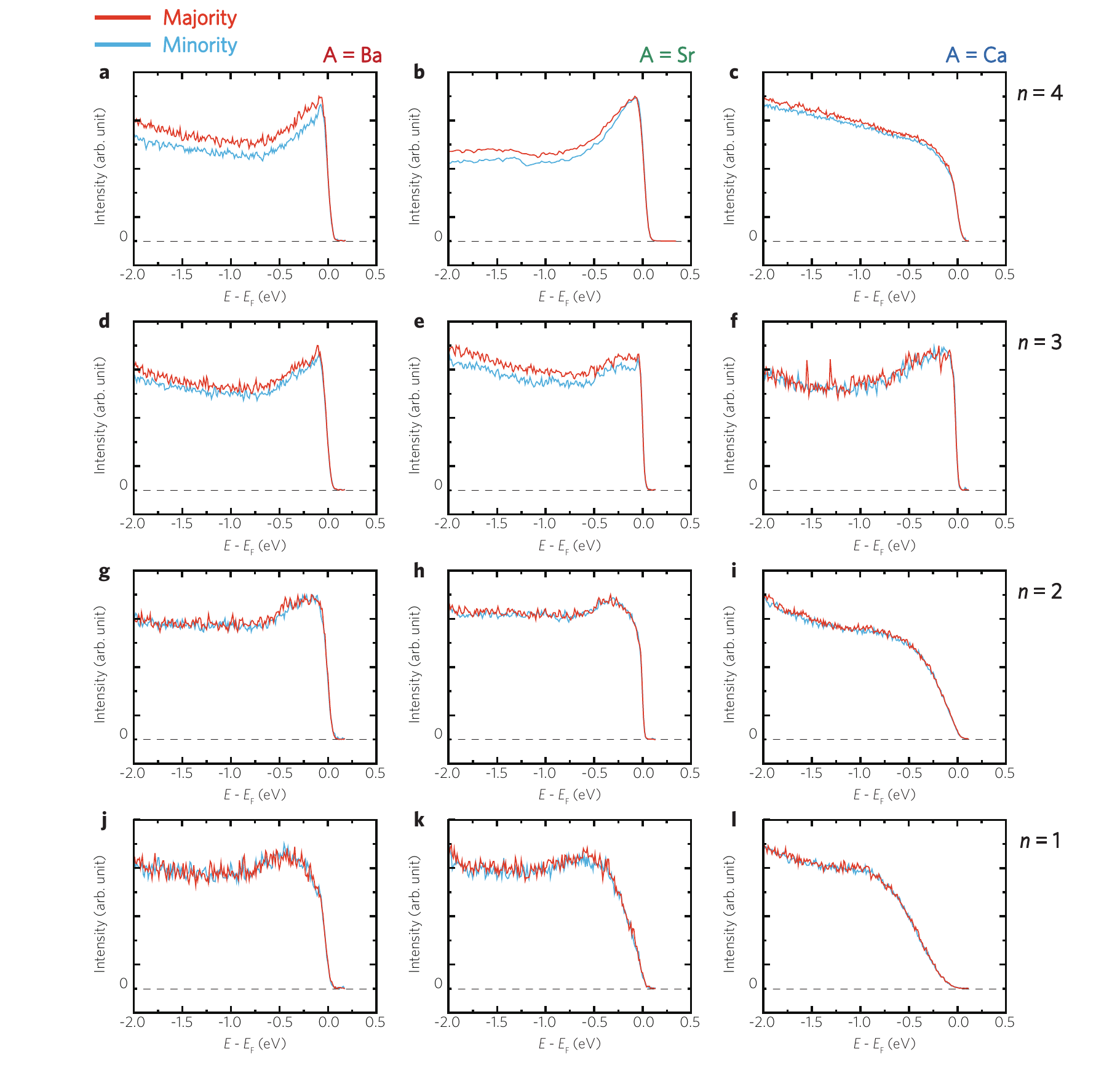}
\caption{{\bf Spin-resolved ARPES measurement of {\it n}A systems ({\it n} $=$ 1--4, A $=$ Ba, Sr, and Ca)}. {\bf a}--{\bf l}, Spin-resolved ARPES measurement of 4Ba ({\bf a}), 4Sr ({\bf b}), 4Ca ({\bf c}), 3Ba ({\bf d}), 3Sr ({\bf e}), 3Ca ({\bf f}), 2Ba ({\bf g}), 2Sr ({\bf h}), 2Ca ({\bf i}), 1Ba ({\bf j}), 1Sr ({\bf k}), and 1Ca ({\bf l}). The majority and minority spins are colored in red and blue, respectively. {\bf b}, {\bf e}, {\bf h}, and {\bf k} were adapted from ref. \cite{sohn2021observation}.}
\label{Ch1}
\end{figure*}

\renewcommand{\thefigure}{S3}
\begin{figure*}[h!]
\centering
\includegraphics[width=\linewidth]{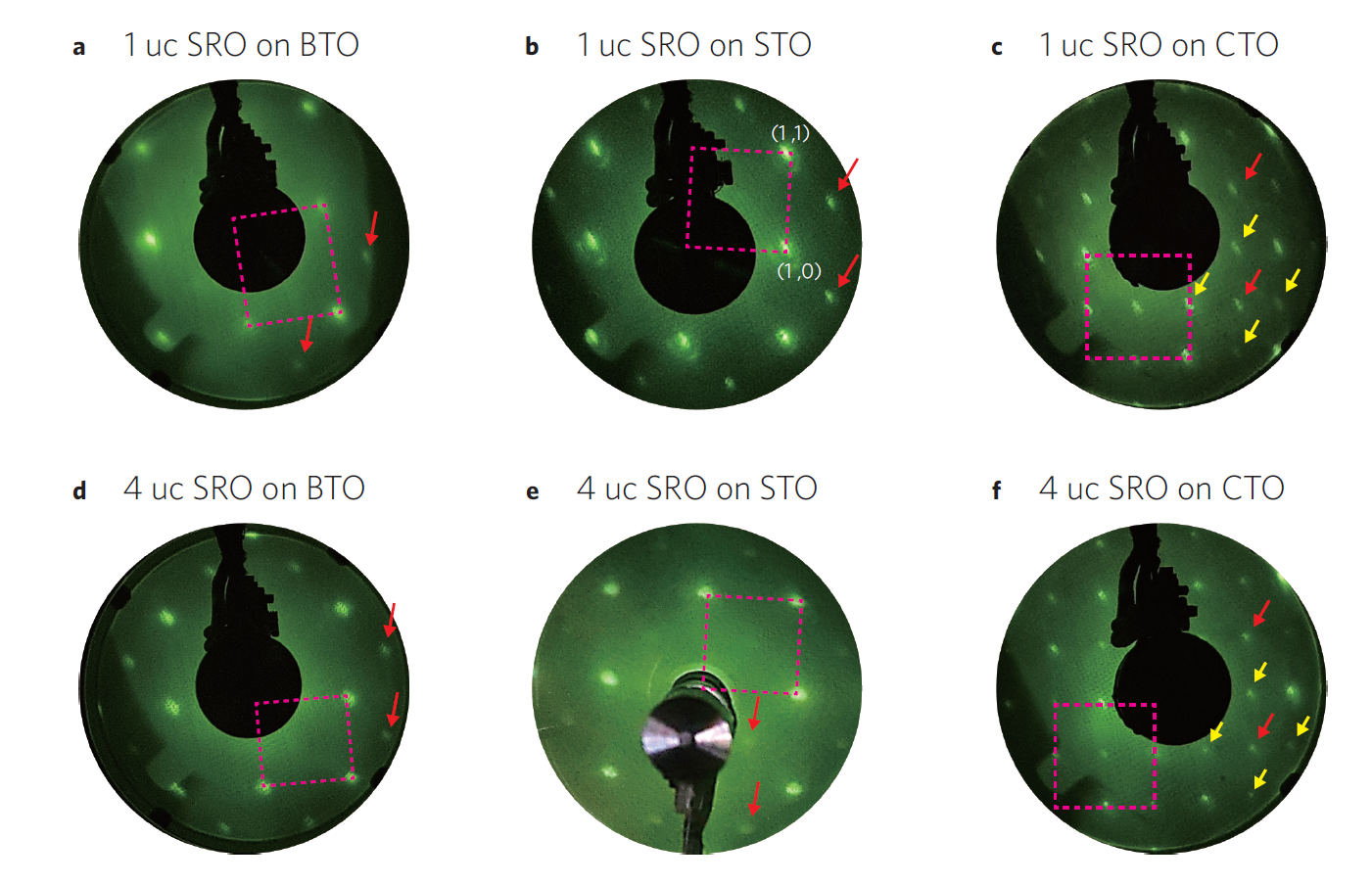}
\caption{{\bf Low-energy electron diffraction (LEED) of the 4A and 1A systems.} {\bf a}--{\bf c}, LEED patterns of 1Ba ({\bf a}), 1Sr ({\bf b}), and 1Ca ({\bf c}). The pink dotted squares show the surface primitive cell of SRO. 1Ba and 1Sr show $\sqrt{2}\times\sqrt{2}$ reconstruction peaks, marked with red arrows. On the other hand, the 1Ca has 2$\times$2 reconstruction marked with red and yellow arrows. Note that the red and yellow arrows represent surface reconstruction peaks corresponding to the $\sqrt{2}\times\sqrt{2}$ and 2$\times$2 reconstructions, respectively. {\bf d}--{\bf f}, LEED patterns of 4Ba ({\bf d}), 4Sr ({\bf e}), and 4Ca ({\bf f}). 4Ba and 4Sr show $\sqrt{2}\times\sqrt{2}$ reconstruction, and 4Ca show 2$\times$2 reconstruction. {\bf b} and {\bf e} were adapted from ref. \cite{sohn2021observation}.}
\label{Ch1}
\end{figure*}

\renewcommand{\thefigure}{S4}
\begin{figure*}[h!]
\centering
\includegraphics[width=\linewidth]{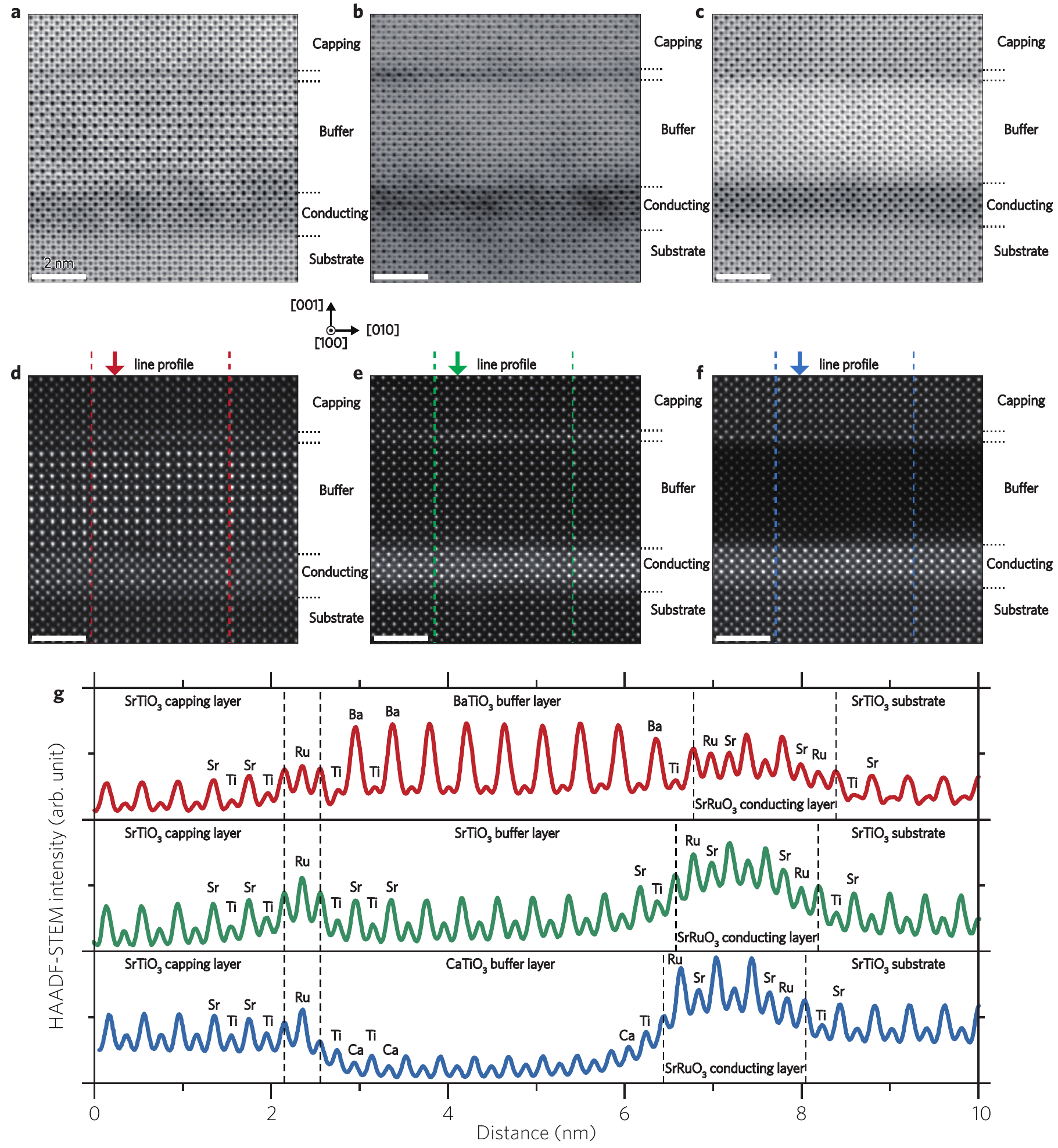}
\caption{{\bf Scanning transmission electron microscopy (STEM) imaging of the 1A systems.} {\bf a}--{\bf c}, Annular bright-field STEM (ABF-STEM) imaging of 1Ba ({\bf a}), 1Sr ({\bf b}), and 1Ca ({\bf c}). {\bf d}--{\bf f}, High-angle annular dark-field STEM (HAADF-STEM) imaging of 1Ba ({\bf d}), 1Sr ({\bf e}), and 1Ca ({\bf f}). All heterostructures are capped with 10-unit-cell (10-UC) STO capping layers. The zone axes are in the [100] direction of the STO substrate, and the white scale bars are 2 nm long. {\bf g}, Line profile of the HAADF-STEM intensity along the [00$\overline{1}$] direction. The HAADF-STEM intensities were integrated inside the red ({\bf d}, 1Ba), green ({\bf e}, 1Sr), and blue ({\bf f}, 1Ca) dotted boundaries.  }
\label{Ch1}
\end{figure*}

\renewcommand{\thefigure}{S5}
\begin{figure*}[h!]
\centering
\includegraphics[width=\linewidth]{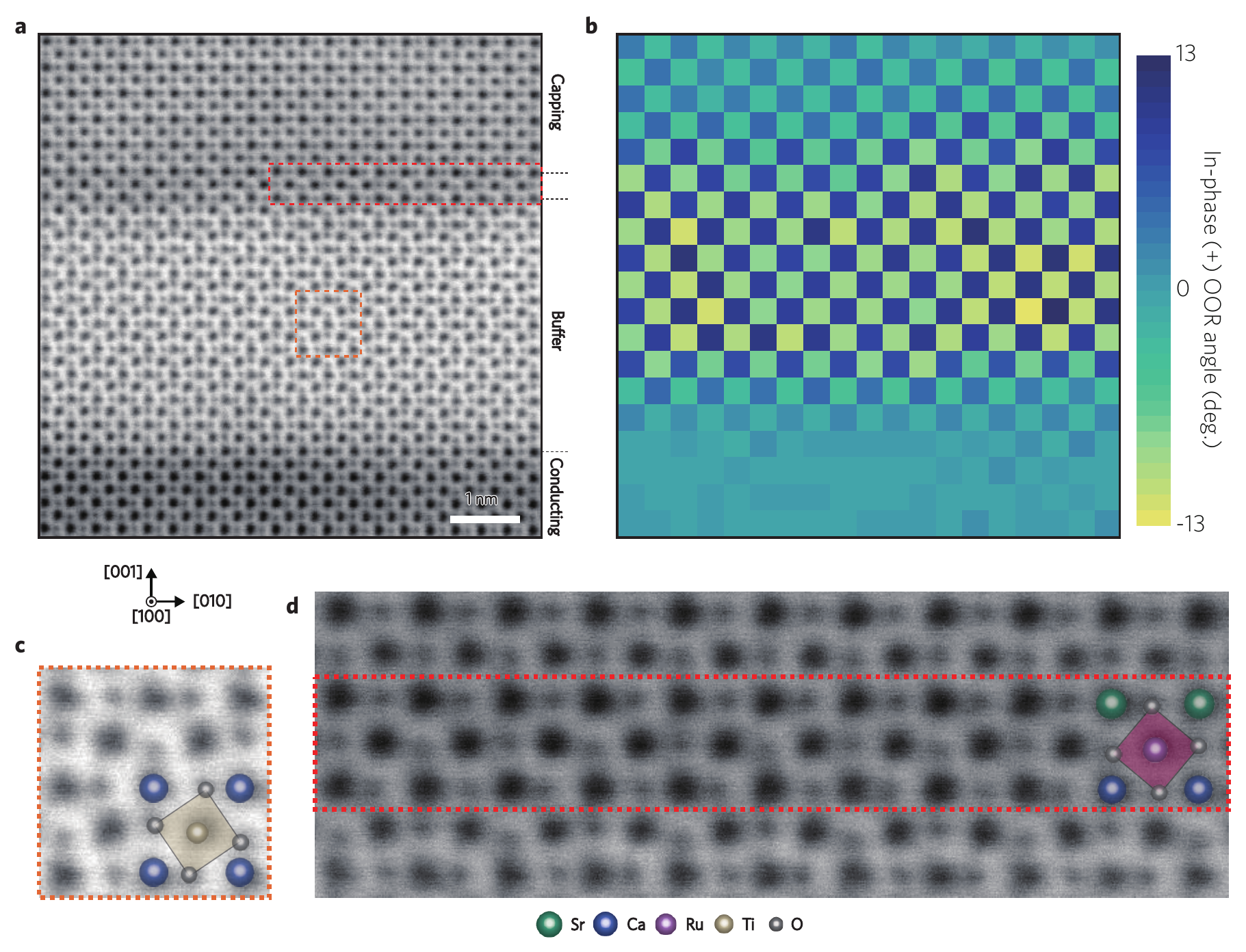}
\caption{{\bf STEM imaging of the in-phase oxygen octahedral rotation (OOR) in the 1Ca.} {\bf a}, ABF-STEM imaging of the 1Ca heterostructure. The zone axis is in the [100] direction of the STO substrate and the white scale bar is 1 nm long. {\bf b}, Mapping of the in-phase OOR of each oxygen octahedron in {\bf a}. {\bf c}, Magnified image of the orange dashed box in {\bf a} that shows the CTO buffer layer. {\bf d}, Magnified image of the red dashed box in {\bf a} that shows the SRO monolayer.}
\label{Ch1}
\end{figure*}

\renewcommand{\thefigure}{S6}
\begin{figure*}[h!]
\centering
\includegraphics[width=\linewidth]{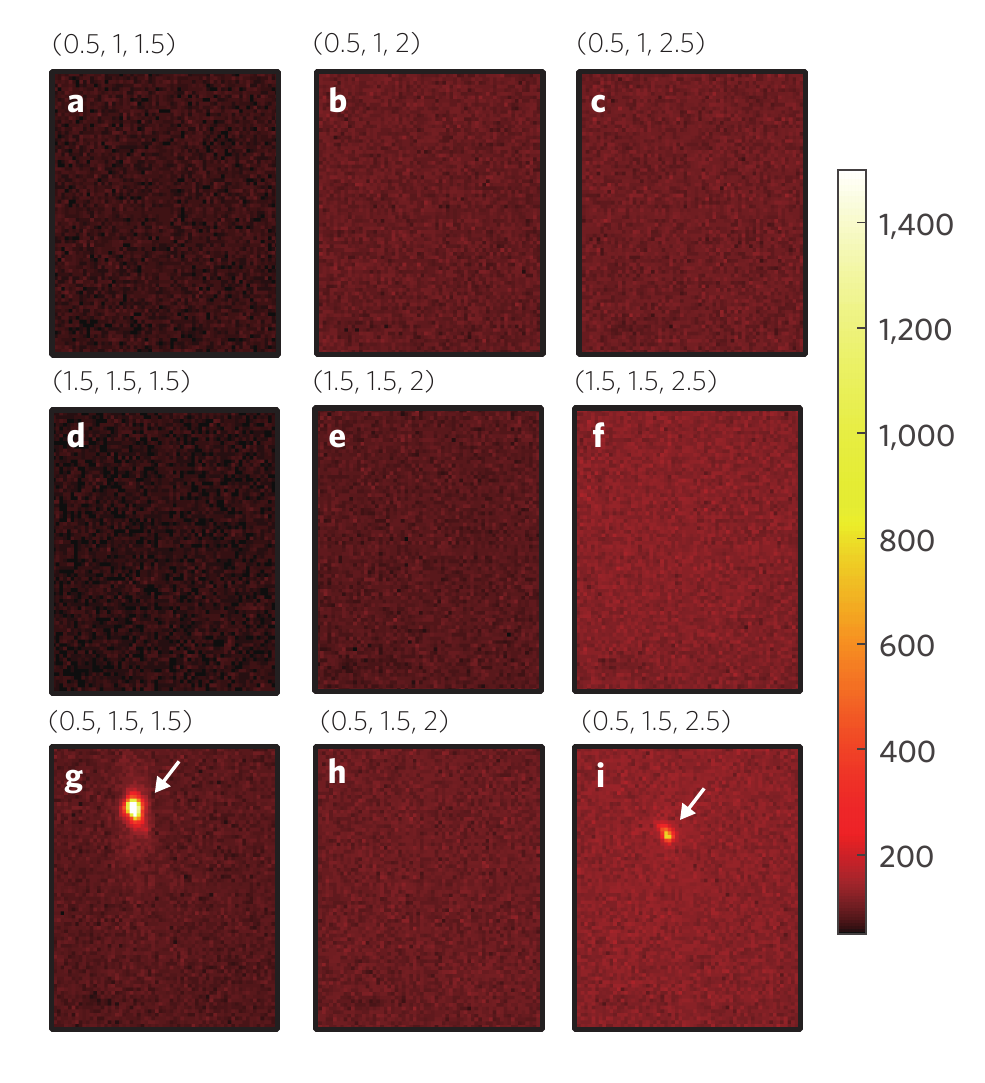}
\caption{{\bf Two-dimensional detector images of the half-order Bragg peaks of 4Ba.} {\bf a}--{\bf c}, Detector images measured along the (0.5 1 {\it L}) rod, where {\it L}~$=$~1.5, 2, and 2.5, respectively. None of the half-order Bragg peaks are observed. {\bf d}--{\bf f}, Detector images measured along the (1.5 1.5 {\it L}) rod, where {\it L}~$=$~1.5, 2 and 2.5, respectively. None of the half-order Bragg peaks are observed. {\bf g}--{\bf i}, Detector images measured along the (0.5 1.5 L) rod, where L~$=$~1.5, 2 and 2.5, respectively. Half-order Bragg peaks are observed when {\it L}~$=$~1.5 and 2.5. All of the observed half-order Bragg peaks ($h/2$ $k/2$ $l/2$), where $h$, $k$, and $l$ are integers, appear when $h$, $k$, and $l$ are odd with $h$~$\neq$~$k$. Considering that the STO substrate and BTO buffer layer do not have any OOR at room temperature, our half-order Bragg peak data suggest that the heterostructure has a tetragonal structure with the $a^0a^0c^-$ OOR \cite{glazer1975simple}.}
\label{Ch1}
\end{figure*}
%%%%%%%%%%%%%%%%%%%%%%%%%%%%%%%%%%%%%%%%%%%%%

\renewcommand{\thefigure}{S7}
\begin{figure*}[h!]
\centering
\includegraphics[width=\linewidth]{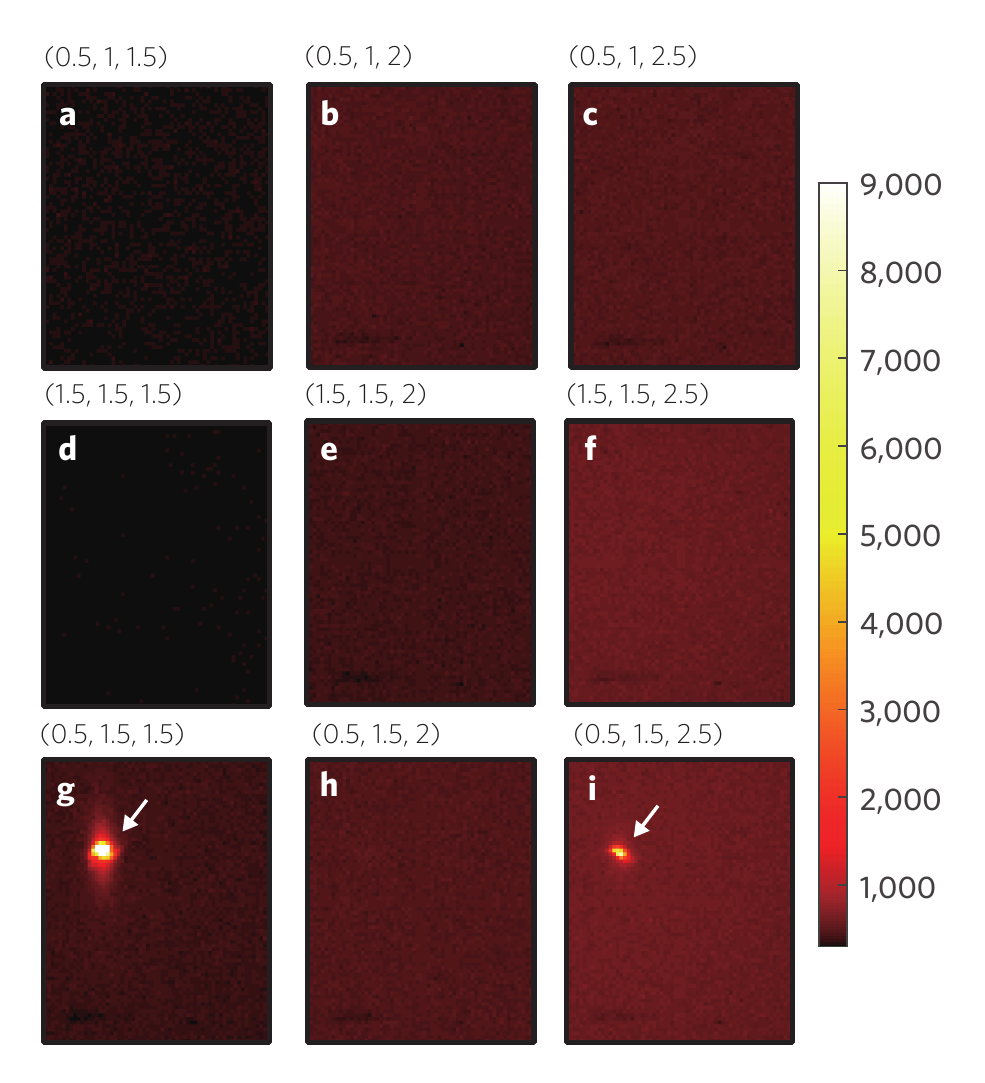}
\caption{{\bf Two-dimensional detector images of the half-order Bragg peaks of 4Sr.} {\bf a}--{\bf c}, Detector images measured along the (0.5 1 {\it L}) rod, where {\it L}~$=$~1.5, 2, and 2.5, respectively. None of the half-order Bragg peaks are observed. {\bf d}--{\bf f}, Detector images measured along the (1.5 1.5 {\it L}) rod, where {\it L}~$=$~1.5, 2 and 2.5, respectively. None of the half-order Bragg peaks are observed. {\bf g}--{\bf i}, Detector images measured along the (0.5 1.5 L) rod, where L~$=$~1.5, 2 and 2.5, respectively. Half-order Bragg peaks are observed when {\it L}~$=$~1.5 and 2.5. All of the observed half-order Bragg peaks ($h/2$ $k/2$ $l/2$), where $h$, $k$, and $l$ are integers, appear when $h$, $k$, and $l$ are odd with $h$~$\neq$~$k$. Considering that the STO substrate has a cubic structure at room temperature, our half-order Bragg peak data suggest that the heterostructure has a tetragonal structure with the $a^0a^0c^-$ OOR \cite{glazer1975simple}.}
\label{Ch2}
\end{figure*}
%%%%%%%%%%%%%%%%%%%%%%%%%%%%%%%%%%%%%%%%%%%%%

\renewcommand{\thefigure}{S8}
\begin{figure*}[h!]
	\centering
	\includegraphics[width=\linewidth]{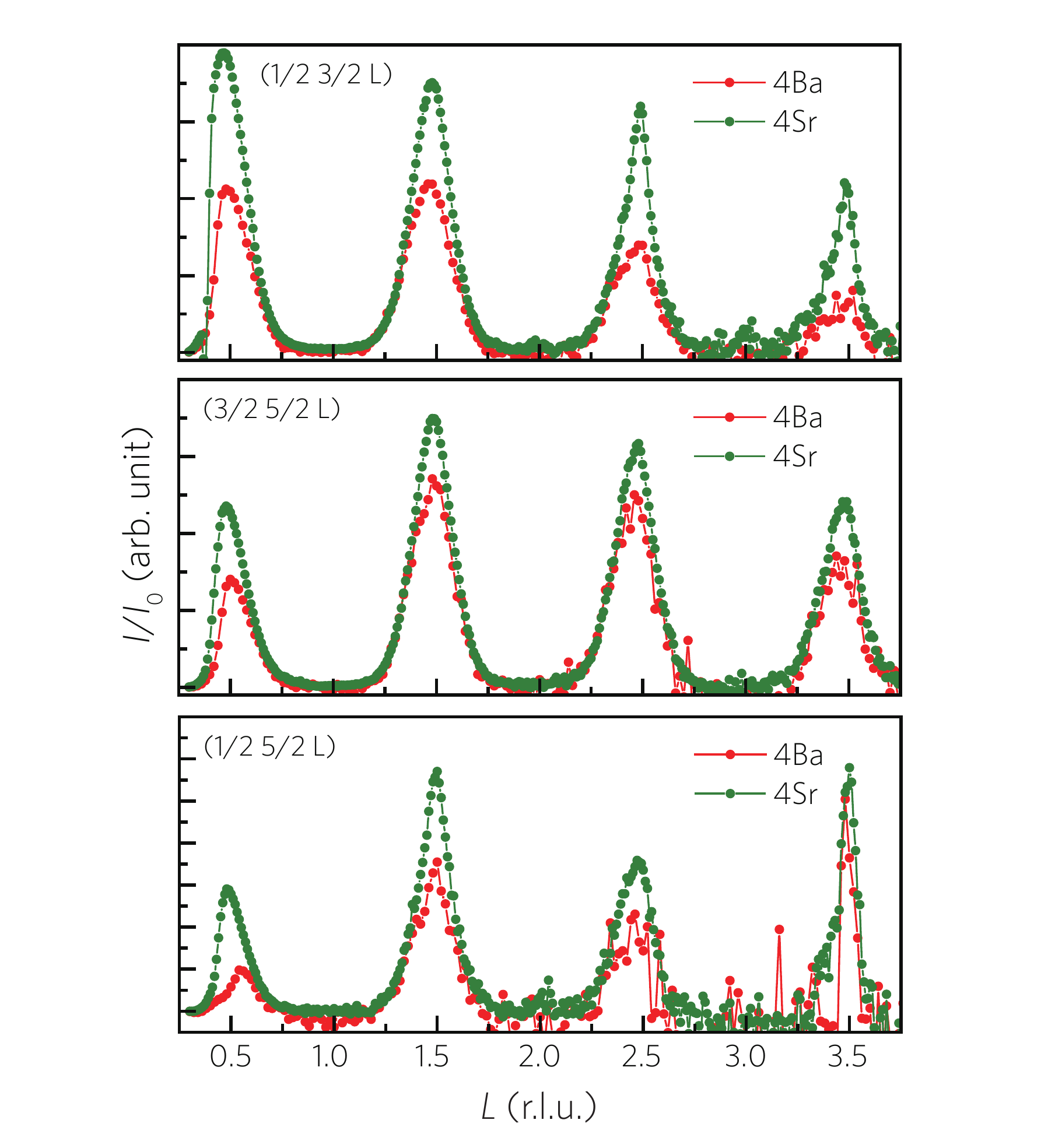}
	\caption{{\bf Half-order Bragg peaks of the modified 4Ba and 4Sr.} The normalized intensities of the half-order Bragg peak measured along the (0.5 1.5 {\it L}), (1.5 2.5 {\it L}), and (0.5 2.5 {\it L}) rods of the modified 4Ba and 4Sr. All half-order Bragg peaks of the modified 4Sr show more strong normalized intensities than those of the modified 4Ba.}
	\label{Ch2}
\end{figure*}
%%%%%%%%%%%%%%%%%%%%%%%%%%%%%%%%%%%%%%%%%%%%%

\renewcommand{\thefigure}{S9}
\begin{figure*}[h!]
\centering
\includegraphics[width=\linewidth]{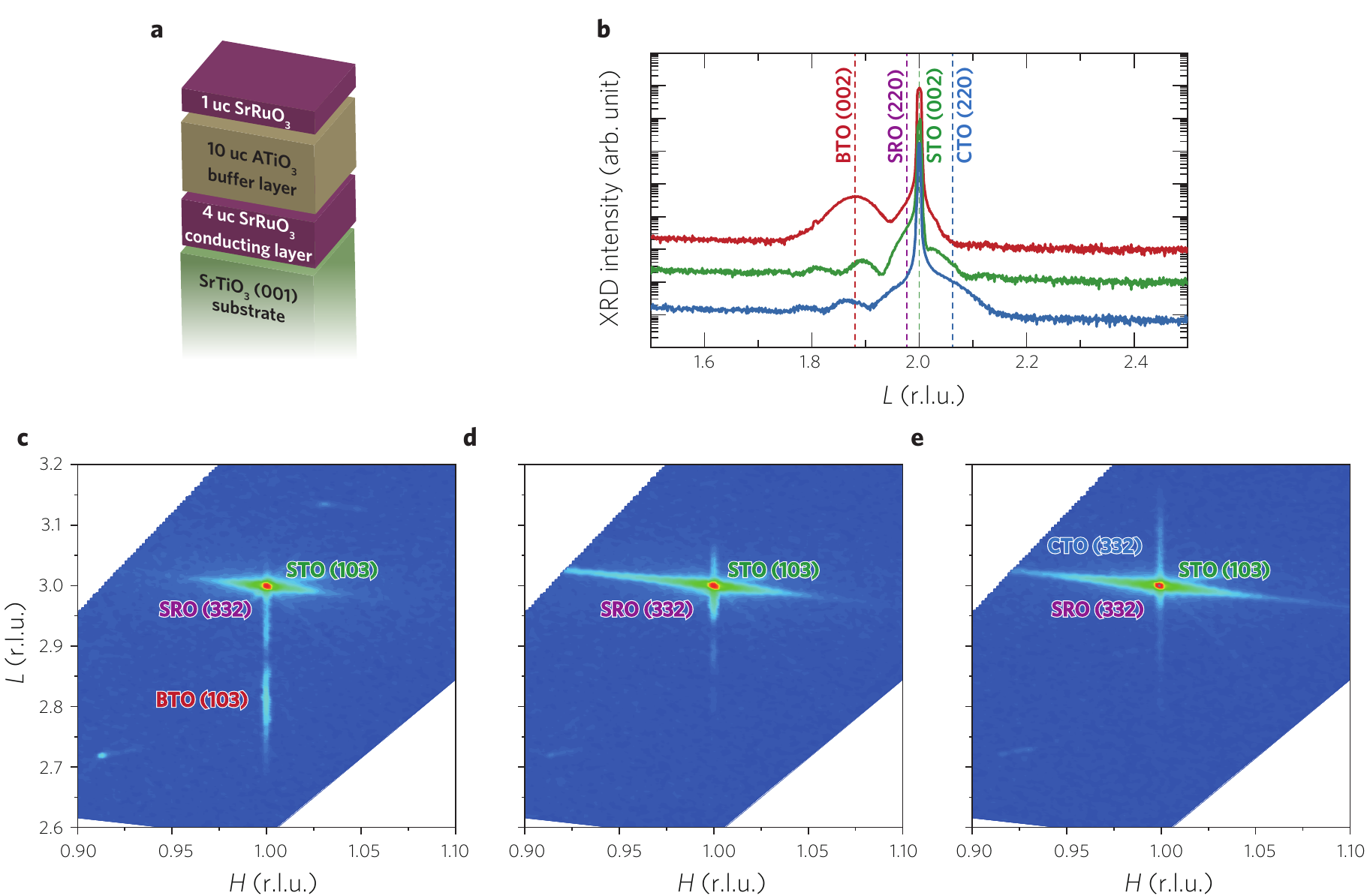}
\caption{{\bf Structural characterization for the 1A systems.} {\bf a}, Schematic of 1A systems composed of a 4-UC SRO layer (conducting layer), 10-UC ATiO$_3$ (ATO, A $=$ Ba, Sr, and Ca) layer (buffer layer), and monolayer SRO, sequentially grown on STO (001) substrates. {\bf b}, XRD 2theta--theta scan for the 1A systems. {\bf c}--{\bf e}, Reciprocal space mapping of the 1A systems around the (103) Bragg peaks of STO substrates.}
\label{Ch1}
\end{figure*}

%%%%%%%%%%%%%%%%%%%%%%%%%%%%%%%%
\renewcommand{\thefigure}{S10}
\begin{figure*}[h!]
\centering
\includegraphics[width=\linewidth]{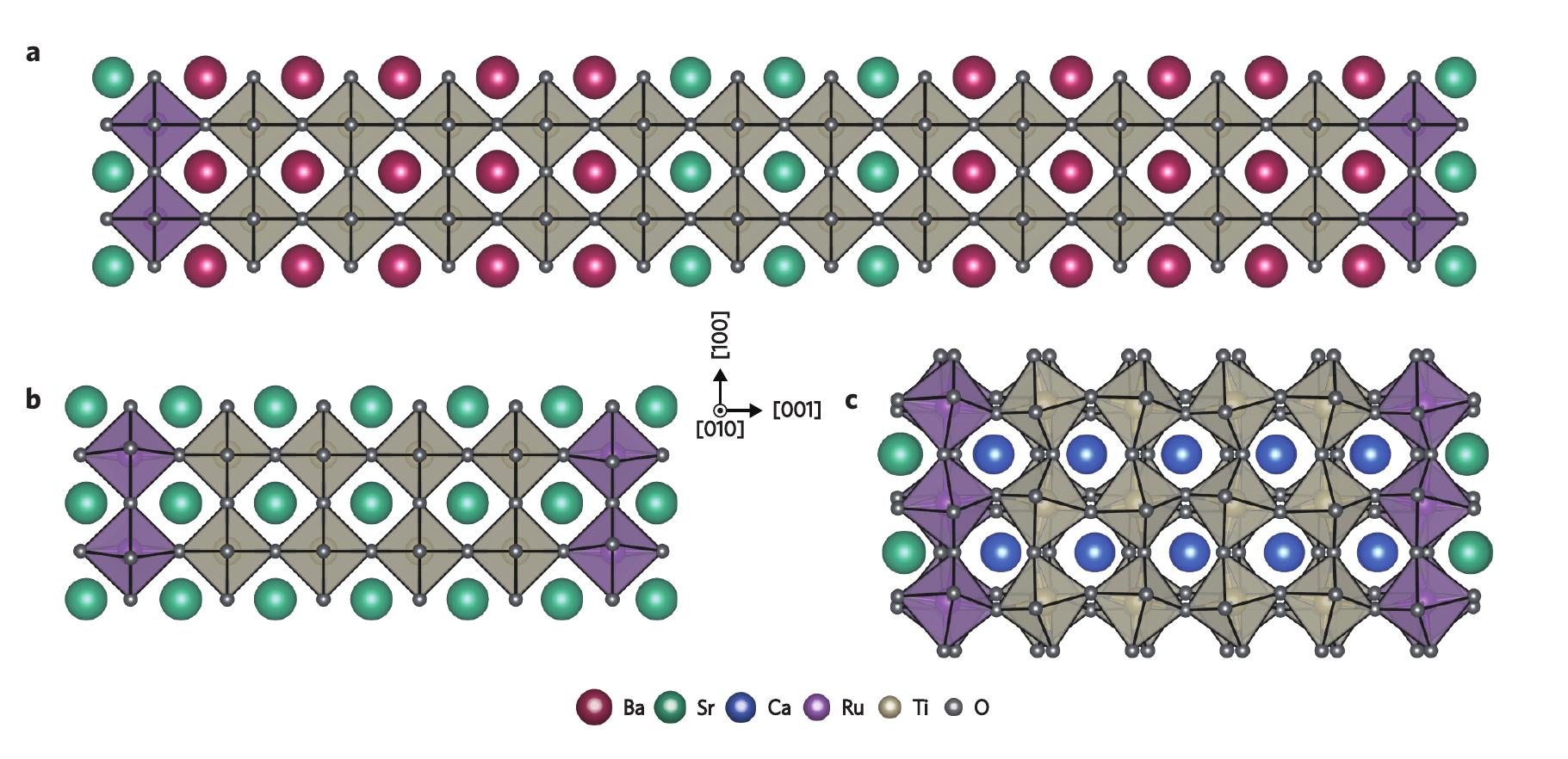}
\caption{{\bf Crystal structure of density functional theory (DFT) calculations.} {\bf a}--{\bf c}, Crystal structure of DFT calculations for 1Ba ({\bf a}), 1Sr ({\bf b}), and 1Ca ({\bf c}).
        }
\label{Ch1}
\end{figure*}

%%%%%%%%%%%%%%%%%

%%%%%%%%%%%%%%%%%%%%%%%%
\renewcommand{\thefigure}{S11}
\begin{figure*}[h!]
\centering
\includegraphics[width=\linewidth]{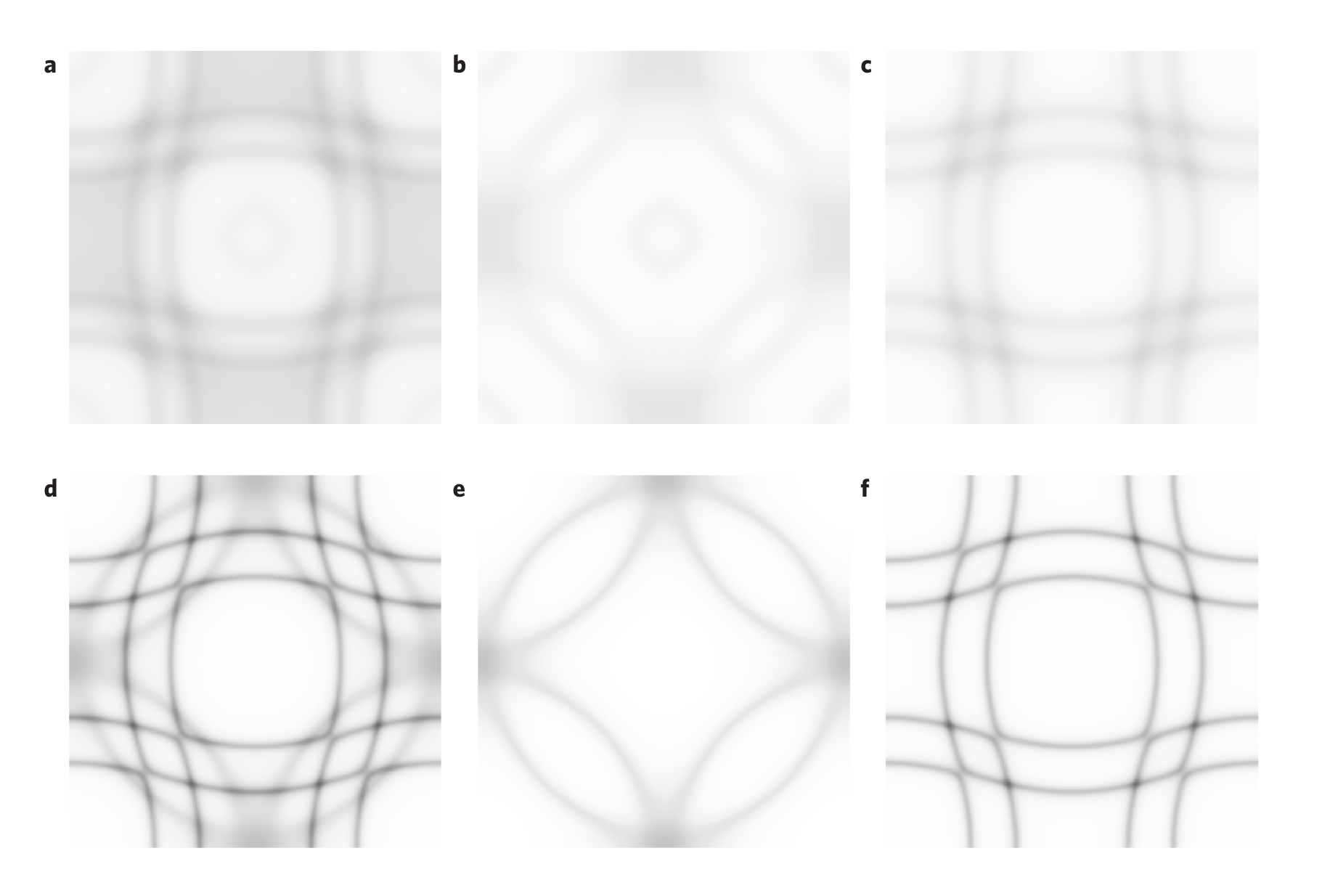}
\caption{\textbf{DFT $+$ dynamical mean-field theory (DMFT) Fermi surfaces.} {\bf a}--{\bf c}, DFT$+$DMFT Fermi surface of 1Sr ({\bf a}) and its projection onto the $d_{xy}$ ({\bf b}) and $d_{yz,zx}$ ({\bf c}) orbitals. {\bf d}--{\bf f}, DFT$+$DMFT Fermi surface of 1Ba ({\bf d}) and its projection onto the $d_{xy}$ ({\bf e}) and $d_{yz,zx}$ ({\bf f}) orbitals.}
\label{Ch1}
\end{figure*}
%%%%

%%%%%
\renewcommand{\thefigure}{S12}
\begin{figure*}[h!]
\centering
\includegraphics[width=\linewidth]{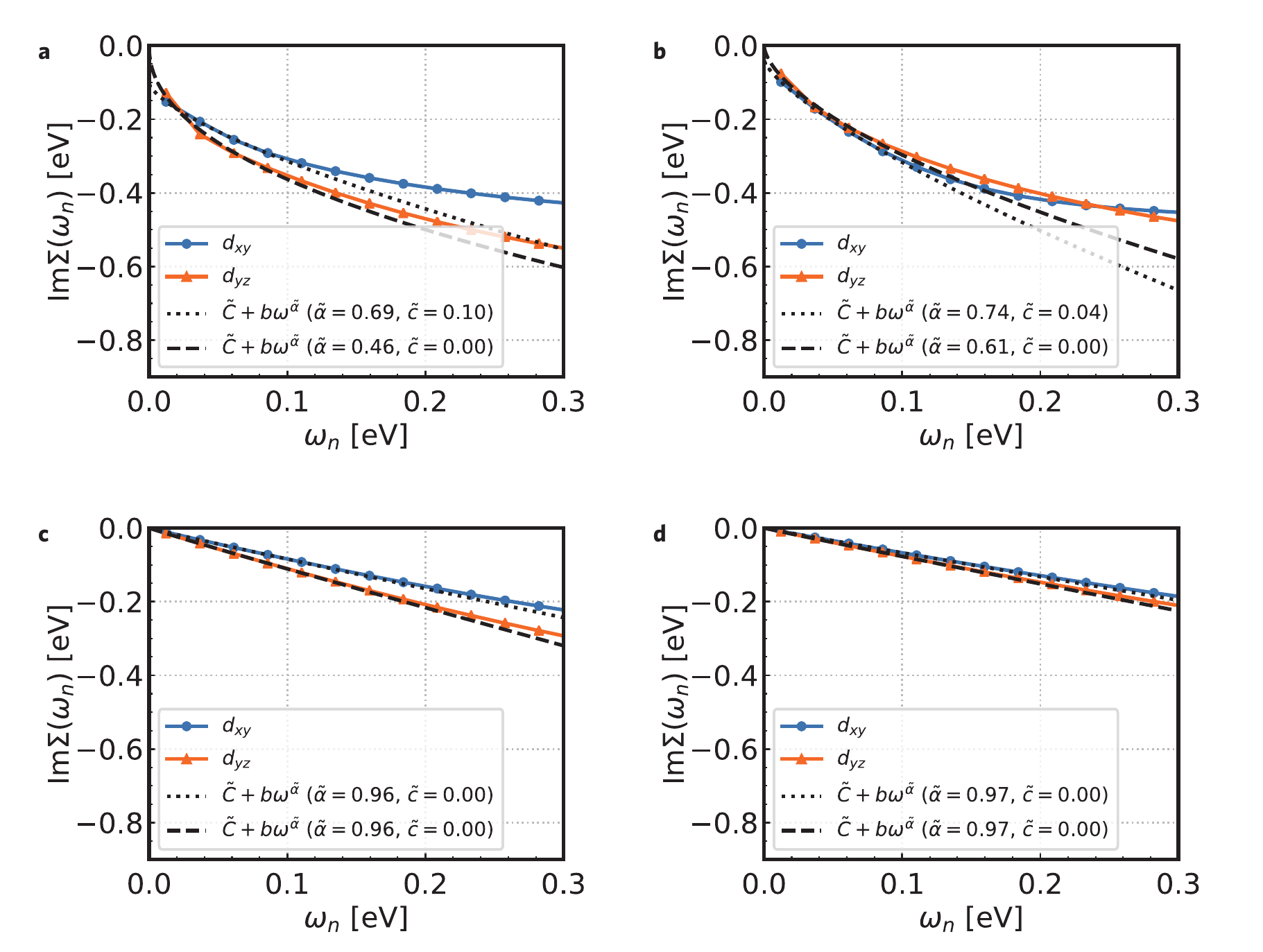}
\caption{{\bf Self-energies in the Matsubara frequency domain.}
		{\bf a},{\bf b}, $d_{xy}$~and $d_{yz}$~components of the Matsubara self-energies and fitting lines for 1Sr ({\bf a}) and 1Ba ({\bf b}) in the presence of Hund coupling ($J_H = 0.4$ eV).
		{\bf c},{\bf d}, $d_{xy}$~and $d_{yz}$~components of the Matsubara self-energies and fitting lines for 1Sr ({\bf c}) and 1Ba ({\bf d}) in the absence of Hund coupling ($J_H = 0.0$ eV).
		}
\label{Ch1}
\end{figure*}
%%%%

%%%%%
\renewcommand{\thefigure}{S13}
\begin{figure*}[h!]
\centering
\includegraphics[width=\linewidth]{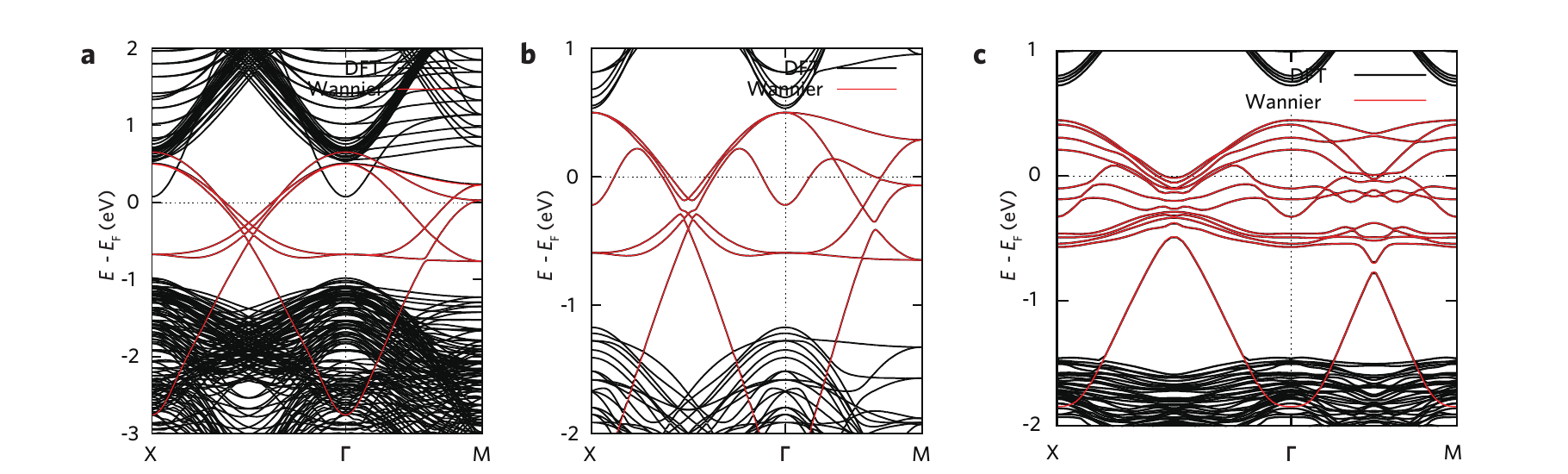}
\caption{\textbf{DFT and maximally localized Wannier function (MLWF) bands.} {\bf a}--{\bf c}, DFT and MLWF bands of 1Ba (\textbf{a}), 1Sr (\textbf{b}), and 1Ca (\textbf{c}).}
\label{Ch1}
\end{figure*}
%%%%

%%%%%
\renewcommand{\thefigure}{S14}
\begin{figure*}[h!]
\centering
\includegraphics[width=\linewidth]{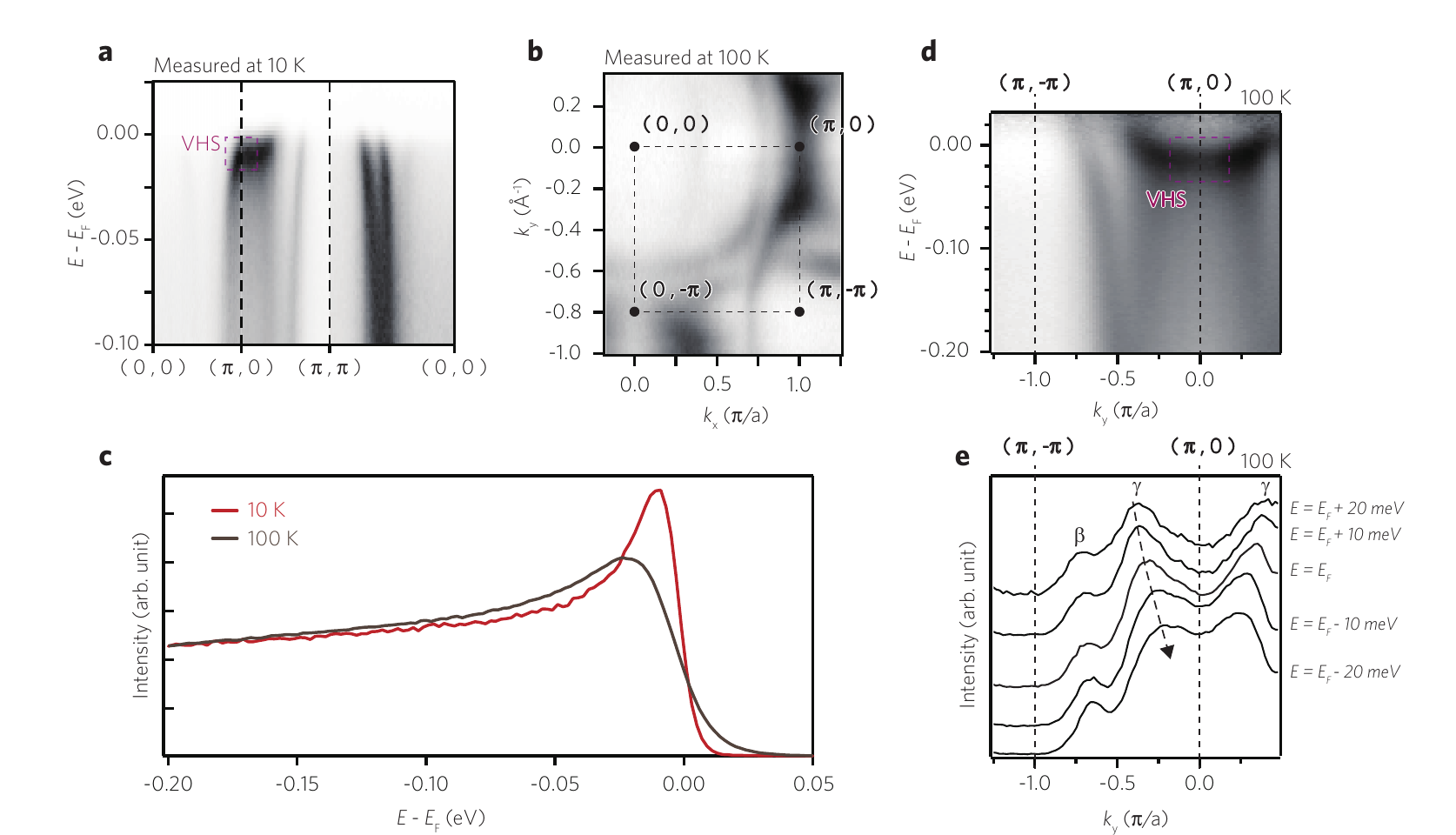}
\caption{{\bf Observation of the van Hove singularity (VHS) in 1Ba.} {\bf a}, High-symmetry cuts of 1Ba measured at 10 K. The VHS of the $\gamma$ band appears below the Fermi level ({\it E}$_F$) at the ( $\pi$ , 0 ) point. {\bf b}, A Fermi surface of 1Ba measured at 100 K. To observe the VHS at the ( $\pi$ , 0 ) point, we performed ARPES at a higher temperature. {\bf c}, Energy distribution curves (EDCs) at the ( $\pi$ , 0 ) point measured at 10 K (red) and 100 K (brown). EDCs are normalized by the intensity at {\it E} $=$ ${\it E}_F-200$ meV. {\bf d}, A high-symmetry cut along the ( $\pi$ , $-\pi$ )--( $\pi$ , 0 ) line in {\bf b}. The high-symmetry cut is divided by the Fermi-Dirac distribution. {\bf e}, Momentum distribution curves near the {\it E}$_F$ extracted from {\bf c}.}
\label{Ch1}
\end{figure*}
%%%%%

%%%%%
\renewcommand{\thefigure}{S15}
\begin{figure*}[h!]
	\centering
	\includegraphics[width=\linewidth]{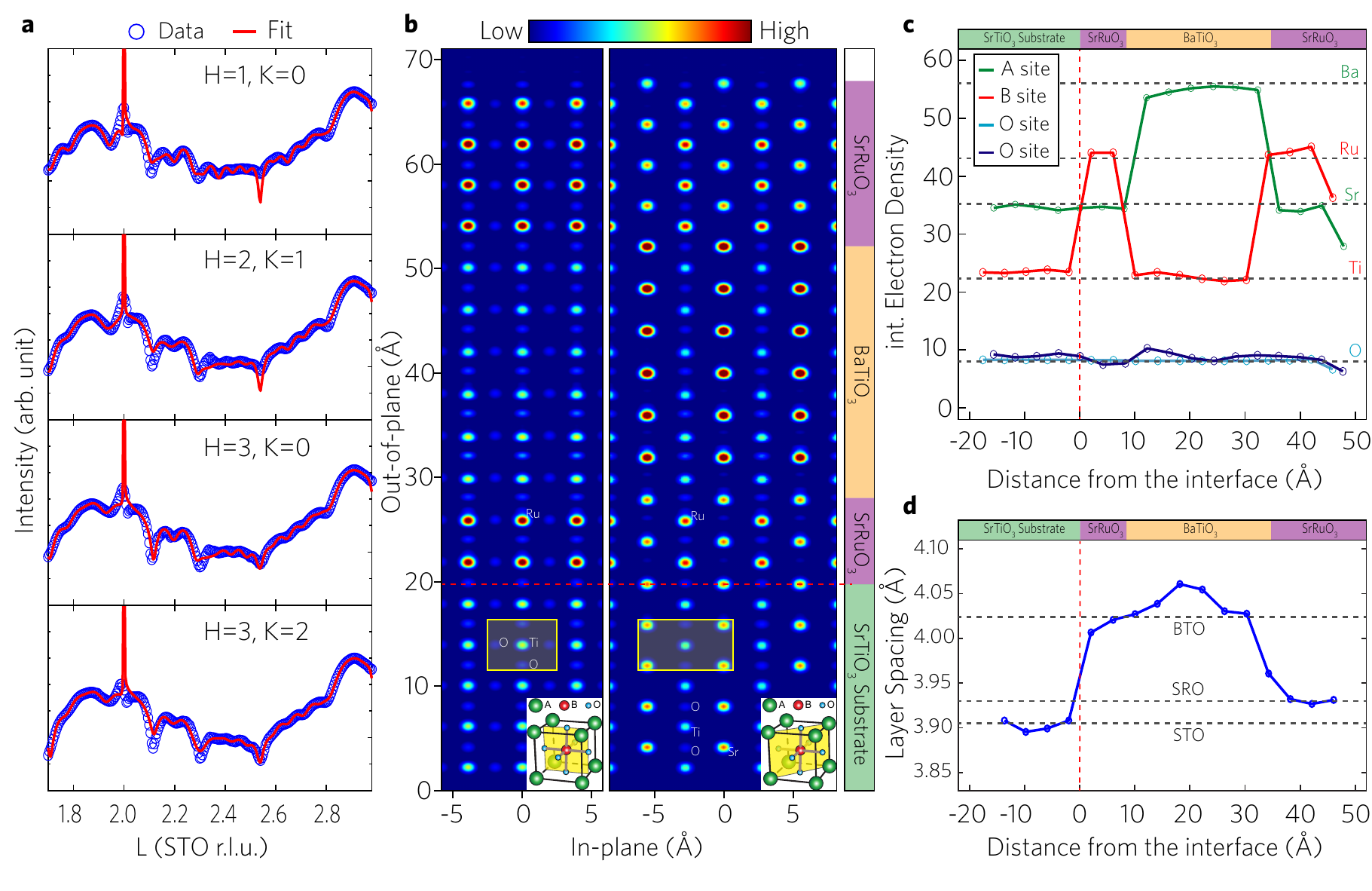}
	\caption{{\bf Coherent Bragg rod analysis of the modified 4Ba.} {\bf a}, Measured crystal truncation rods (blue circles) and fits (red lines) for the modified 4Ba. The indices in each plot refer to the (h, k, l) of the SrTiO$_3$ substrate measured for each CTR.
	{\bf b}, Vertical cuts through a COBRA-reconstructed three-dimensional (3D) electron density map along the [100] and [110] directions for the modified 4Ba. Insets show schematic diagrams of specific plane cuts (yellow) with respect to the perovskite unit cell depicted in each 2D electron density map. The yellow square indicates the cubic/pseudo-cubic unit cell, and the red dotted line denotes the interface between the substrate and film.
	{\bf c}, Profile of the integrated electron density for each perovskite site as a function of the distance from the substrate-film interface. Nominal electron densities for Ba, Ru, Sr, Ti, and O atoms at an incident photon energy of 15.5~keV are shown as horizontal dashed lines. 
	{\bf d}, Layer-dependent out-of-plane lattice spacing determined from the distance between A- and B-site cations. The bulk out-of-plane lattice parameters for SrTiO$_3$, SrRuO$_3$, and BaTiO$_3$ are shown as horizontal dashed lines.}
	\label{Ch2}
\end{figure*}
%%%%%%%%%%%%%%%%%%%%%%%%%%%%%%%%%%%%%%%%%%%%%

%%%%%
\renewcommand{\thefigure}{S16}
\begin{figure*}[h!]
	\centering
	\includegraphics[width=\linewidth]{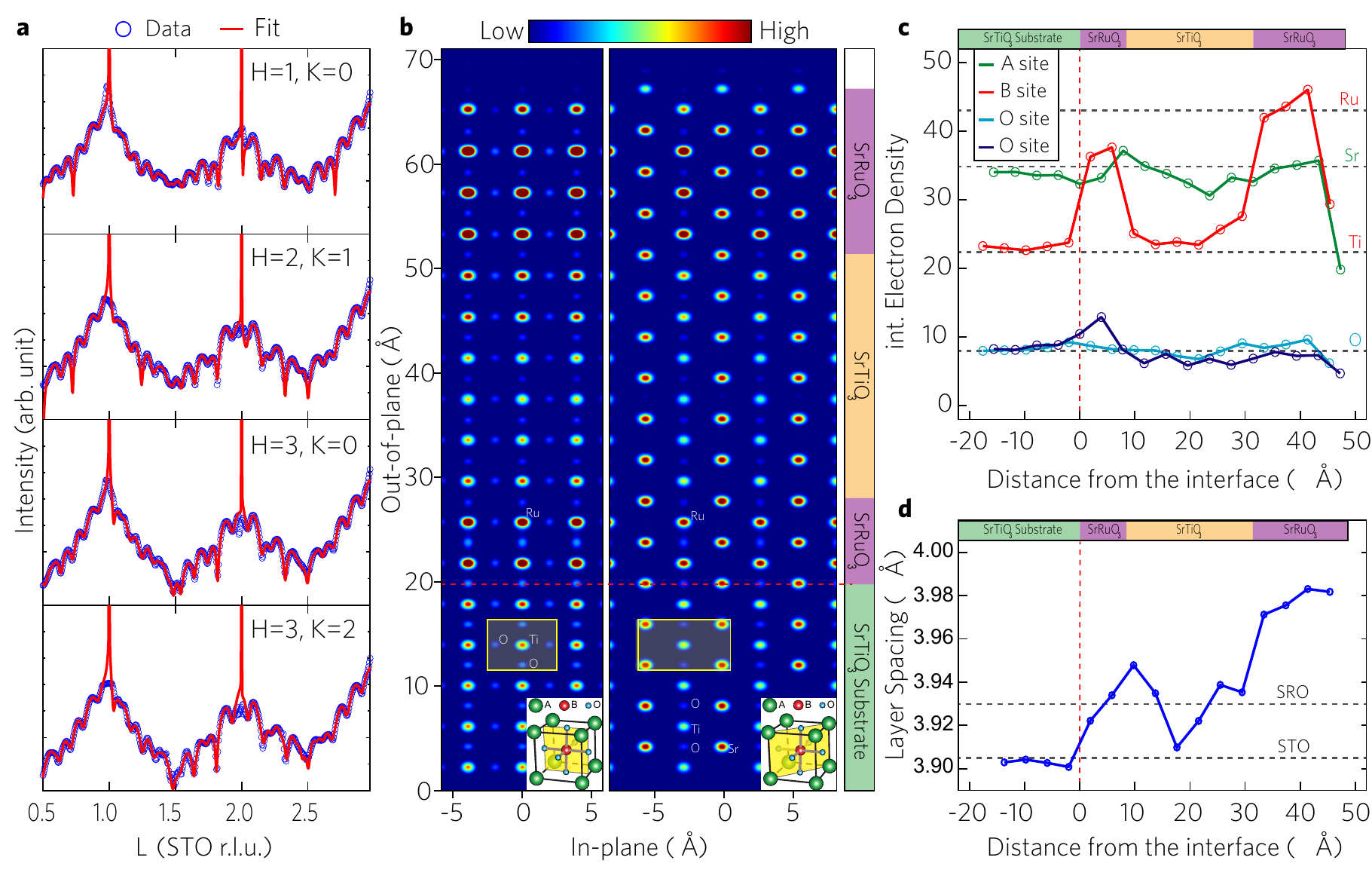}
	\caption{{\bf Coherent Bragg rod analysis of the modified 4Sr.} {\bf a}, Measured crystal truncation rods (blue circles) and fits (red lines) for the modified 4Sr. The indices in each plot refer to the (h, k, l) of the SrTiO$_3$ substrate measured for each CTR.
	{\bf b}, Vertical cuts through a COBRA-reconstructed 3D electron density map along the [100] and [110] directions for the modified 4Sr. Insets show schematic diagrams of specific plane cuts (yellow) with respect to the perovskite unit cell depicted in each 2D electron density map. The yellow square indicates the cubic/pseudo-cubic unit cell, and the red dotted line denotes the interface between the substrate and film.
	{\bf c}, Profile of the integrated electron density for each perovskite site as a function of the distance from the substrate-film interface. Nominal electron densities for Ru, Sr, Ti, and O atoms at an incident photon energy of 15.5~keV are shown as horizontal dashed lines. 
	{\bf d}, Layer-dependent out-of-plane lattice spacing determined from the distance between A- and B-site cations. The bulk out-of-plane lattice parameters for SrTiO$_3$ and SrRuO$_3$ are shown as horizontal dashed lines.
	}
	\label{Ch2}
\end{figure*}
%%%%%%%%%%%%%%%%%%%%%%%%%%%%%%%%%%%%%%%%%%%%%
%%%%%

\end{document}